\documentclass[journal,a4paper]{IEEEtran}

\usepackage{cite}
\usepackage{array}
\usepackage{fixltx2e}
\usepackage{color}
\usepackage{psfrag}
\usepackage{epsfig}
\usepackage{tabularx}
\usepackage{amsmath}
\usepackage{amssymb}
\usepackage{amsfonts}
\usepackage{algorithm,algorithmic}
\usepackage{pstricks, pst-node, pst-plot, pst-circ}
\usepackage{moredefs}
\usepackage{algorithm,algorithmic}

\def\PSNR{\mathrm{ PSNR}}

\def\ve#1{{\mathchoice{\mbox{\boldmath$\displaystyle #1$}}%
		      {\mbox{\boldmath$\textstyle #1$}}%
		      {\mbox{\boldmath$\scriptstyle #1$}}%
		      {\mbox{\boldmath$\scriptscriptstyle #1$}}}} 

\def\diag{\mathrm{ diag}}

\def\diag{\mathrm{ diag}}

\def\e{\mathrm{ e}}

\def\j{\mathrm{ j}}

\def\trans{\mathsf{T}}
\def\herm{\mathsf{H}}

\def\dB{\,\mathrm{ dB}}

\newcommand{\argmin}{\mathop{\mathrm{argmin}}}
\newcommand{\argmax}{\mathop{\mathrm{argmax}}}

\newcolumntype{C}[1]{>{\centering\arraybackslash}p{#1}}

\newcommand{\algorithmicinput}{\textbf{input:}}
\newcommand{\INPUT}{\item[\algorithmicinput]}
\newcommand{\algorithmicoutput}{\textbf{output:}}
\newcommand{\OUTPUT}{\item[\algorithmicoutput]}

\begin{document}

	
\title{Increasing Imaging Resolution by Non-Regular Sampling and Joint Sparse Deconvolution and Extrapolation}
\author{J{\"u}rgen~Seiler, \emph{IEEE Senior Member}, Markus Jonscher, Thomas Ussmueller, \emph{IEEE Senior Member},\\ and~Andr{\'e}~Kaup, \emph{IEEE Fellow}
\thanks{J.\ Seiler, M. Jonscher, and A.\ Kaup are with the Chair of Multimedia Communications and Signal Processing, Friedrich-Alexander-Universit{\"a}t Erlangen-N{\"u}rnberg (FAU), Cauerstr. 7, 91058 Erlangen, Germany (e-mail: juergen.seiler@FAU.de; \mbox{markus.jonscher@FAU.de}; \mbox{andre.kaup@FAU.de}). T. Ussmueller is with the Institute for Mechatronics, Leopold-Franzens-Universit{\"a}t Innsbruck,  Innrain 52, 6020 Innsbruck, Austria (e-mail: thomas.ussmueller@uibk.ac.at). 

} \vspace{-0.5cm}}

\markboth{}%
{Seiler, Jonscher, Ussmueller, Kaup: Three-Quarter Sampling}

\maketitle


\begin{abstract} \label{abstract}
Increasing the resolution of image sensors has been a never ending struggle since many years. In this paper, we propose a novel image sensor layout which allows for the acquisition of images at a higher resolution and improved quality. For this, the image sensor makes use of non-regular sampling which reduces the impact of aliasing. Therewith, it allows for capturing details which would not be possible with state-of-the-art sensors of the same number of pixels. The non-regular sampling is achieved by rotating prototype pixel cells in a non-regular fashion. As not the whole area of the pixel cell is sensitive to light, a non-regular spatial integration of the incident light is obtained. Based on the sensor output data, a high-resolution image can be reconstructed by performing a deconvolution with respect to the integration area and an extrapolation of the information to the insensitive regions of the pixels. To solve this challenging task, we introduce a novel joint sparse deconvolution and extrapolation algorithm. The union of non-regular sampling and the proposed reconstruction allows for achieving a higher resolution and therewith an improved imaging quality.

\end{abstract}


\section{Introduction} \label{sec:introduction}
Looking at the development of camera systems in the past years, an ongoing pursuit for higher resolutions can be discovered. Up to now, this has mainly been achieved by increasing the number of pixels used in the image sensor inside the camera. That is to say, more and more light-sensitive elements are used for obtaining a high-resolution sensor and therewith allowing for high-resolution imaging. However, this practice has the disadvantage that an increase of the number of pixels typically comes along with a higher price and a larger power consumption. Furthermore, a reasonable reduction of the size of the individual pixels is only possible up to a certain point due to photometric limits~\cite{Bruckner2013}. Hence, if one wants to avoid increasing the actual sensor dimensions, the number of pixels per area is limited.

One way for increasing the resolution of images is to estimate high-resolution information from the acquired image by post-processing, even if the underlying sensor is not able to theoretically resolve such fine details. All these post-processing operations belong to the group of super-resolution (SR) techniques. Accordingly, SR techniques can be divided into three groups. First, this is the classical multi-image SR~\cite{Park2003}. The algorithms from this group exploit that multiple images of the same object or scene are taken consecutively. Based on small displacements caused by movement, a higher resolution can be recovered. However, these techniques are not applicable if only a single image is available. The same holds for the second group of SR techniques which is multi-camera SR~\cite{Richter2016}. There, the information from multiple cameras at different positions can be used for increasing the resolution. Multi-camera SR is able to achieve a very high quality but apparently, it can only be applied if more than one camera is available. Unlike the former two groups which rely either on multiple images in time or in spatial direction, the third group can also work on single images. Thus they are called single-image SR~\cite{Shreyaskumar2012} algorithms. These algorithms exploit properties like self-similarity of the acquired object or scene~\cite{Glasner2009} or use information from training data sets~\cite{Yang2010, Kim2010a, Zeyde2010} for estimating a high-resolution image. In the case that these properties are fulfilled, single-image SR algorithms are able to achieve a very high quality. However, if the underlying assumptions are not met, they fail. All SR concepts have in common that they all rely on the quality of the actual output of existing image sensors and the therewith along-coming limitations. Thus, we are actually aiming at increasing the resolution of the image sensor directly, instead of applying post-processing SR techniques.

Common image sensors are designed on the two-dimensional repetition of a prototype pixel cell. Hence, these sensors perform a regular two-dimensional sampling and due to this the resolution is limited by aliasing. Thus, for increasing the achievable resolution, the image sensor has to be modified in order to reduce the influence of aliasing. For this, we have proposed a slightly modified custom image sensor in~\cite{Schoeberl2011} which allows for a higher imaging quality if used in combination with an appropriate reconstruction algorithm. The modification of the sensor consists of a mask which is put on the image sensor and non-regularly shields three quarters of every pixel. Using this, the acquisition is carried out effectively on a grid twice as fine as the underlying sensor. However, this is achieved at the cost that three quarters of the pixels with respect to the fine grid are missing. Nevertheless, using the reconstruction proposed in~\cite{Schoeberl2011} or the improved Frequency Selective Reconstruction~\cite{Seiler2015}, a very high reconstruction quality can be achieved. However, this technique has the drawback that by masking three quarters of every pixel, only one quarter of the sensor area remains sensitive to light, therewith lowering the overall sensitivity.

Even though the non-regular shielding allows for a higher resolution, sacrifying so much of the sensitivity is not acceptable for practical image sensors. Looking at state-of-the-art image sensors, it becomes obvious that one always tries to make most of the area sensitive to light and the objective hence is to bring the fill-factor close to one. This can be achieved for example by applying techniques like backside illumination~\cite{Blouke1978, Etoh2009} or the use of microlenses~\cite{Ishihara1983, Popovic1988, Catrysse2002, Bruna2010, Jianhua2012}. However, these techniques again lead to a regular sensing and therewith the resolution is limited by aliasing. 

Nevertheless, the influence of aliasing can be reduced by non-regular sampling. As shown in~\cite{Hennenfent2007} and~\cite{Maeda2009}, a non-regular or a random sampling can be used for reducing the visible influence of aliasing. However, this is not limited to visual features as we have shown in ~\cite{Seiler2015}. As discussed there, a higher resolution can be achieved by applying non-regular sampling since the aliasing does not lead to a repetition of the spectral components but rather to a noise-like floor in the spectrum. And this floor can be suppressed by an appropriate reconstruction using a-priori signal knowledge. 

In this paper, we want to propose a new image sensor layout scheme which allows for a non-regular sampling but at the same time does not decrease the sensitivity as significantly as~\cite{Schoeberl2011}. The proposed scheme does not require a new technology, but rather the sensors can be designed using existing tool chains and manufacturing processes. The new layout results from non-regularly rotating a prototype pixel cell and therewith obtaining a non-regular orientation of the light-sensitive areas. As the non-regular placement does not directly lead to a higher resolution, we also propose a novel reconstruction algorithm which is called Joint Sparse Deconvolution and Extrapolation (JSDE). By using the combination of a non-regularly sampling sensor and this fitting reconstruction algorithm, images with four times the resolution of the underlying sensor can be reconstructed.

The paper is structured as follows. In the next section, the proposed sensor layout is discussed in detail and it is shown how a non-regular sampling sensor can be derived from a prototype pixel cell. In Section~\ref{sec:reconstruction}, the novel JSDE algorithm is outlined and it is shown how this algorithm can recover a high-resolution image from the output of the non-regular sampling sensor. As the algorithm can also be interpreted in the Compressed Sensing (CS) framework~\cite{Candes2006, Donoho2006}, the section also contains a brief discussion about the relationship between JSDE and CS algorithms. Afterwards, in Section~\ref{sec:simulations} simulation results are provided in order to show the effectiveness of the combination of a non-regularly sampling sensor together with an appropriate reconstruction. This section also provides a comparison to alternative acquisition concepts and reconstruction algorithms. Finally, a conclusion is provided in \mbox{Section~\ref{sec:conclusion}} together with a short outlook to further developments.


\section{Proposed Sensor Layout} \label{sec:sensorlayout} 
\begin{figure}
	\centering
		\includegraphics[width=0.26\textwidth]{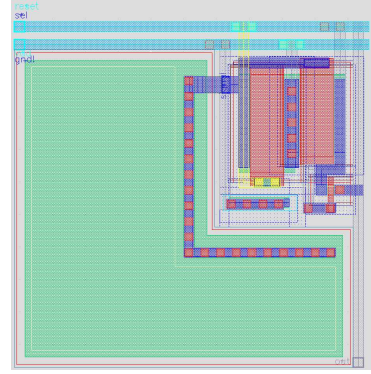}
	\caption{Example of a prototype pixel cell. The top right quadrant contains the circuitry. The other three quadrants form the light-sensitive photo diode.}
	\label{fig:pixel_cell}
\end{figure}

Looking at image sensors, it can be observed that they are typically designed on the basis of a prototype pixel cell, as shown exemplary in Figure~\ref{fig:pixel_cell}, that is repeated many times in horizontal and vertical direction. Accordingly, the light-sensitive pixels are placed on a regular two-dimensional grid. In this context, it always has to be kept in mind that the individual pixels are not sensitive to light over their whole area but also contain insensitive regions resulting from the circuitry within each pixel. Nevertheless, by using techniques like backside illumination~\cite{Blouke1978, Etoh2009} and / or microlenses~\cite{Ishihara1983, Popovic1988, Catrysse2002, Bruna2010, Jianhua2012} a fill-factor of nearly $100 \%$ can be achieved. Using this, one gets an array of rectangular light sensitive areas and the acquisition process can be regarded as an integration of the incident light on identical areas located on the two-dimensional regular grid. 

This acquisition process limits the resolution of imaging sensors in two ways. First, the integration over the light-sensitive areas of every pixel inherently produces a low-pass characteristic, therewith attenuating high spatial frequencies and image details. Second, and more severe, the other limiting factor is the aliasing from the sampling process. The regular placement of the pixels leads to classical aliasing. Due to this, high spatial frequencies get mapped to low-frequency ones, distorting the image quality strongly. In order to avoid this, either a special anti-alias filter has to be placed in front of the image sensor or the used lens has to suppress high frequencies in such a way that the impact of aliasing is small.

In order to resolve this and to be able to achieve a higher resolution with image sensors, we propose a novel sensor layout. This new layout in combination with the proposed joint sparse deconvolution and extrapolation reconstruction algorithm explained in the next section yields a higher image quality. The basic idea of the novel layout is to use a non-regular placement of the light-sensitive areas. As we have shown in~\cite{Seiler2015}, non-regular sampling has the advantage that the aliasing does not lead to a repetition of the spectrum, or respectively, to a mapping of high frequencies to low ones. Instead, non-regular sampling produces a noise-like floor in the frequency domain. Using a sparsity-based reconstruction algorithm, this noise-like floor can be suppressed by exploiting the sparsity property of image signals~\cite{Candes2007}. 

\begin{figure}
	\centering
	\psfrag{GND}[r][r][0.6]{GND}
	\psfrag{VDD}[l][l][0.6]{VDD}
	\psfrag{RESET}[r][r][0.6]{RESET}
	\psfrag{OUT}[r][r][0.6]{OUT}
	\psfrag{ENABLE}[r][r][0.6]{ENABLE}
	\includegraphics[width=0.41\textwidth]{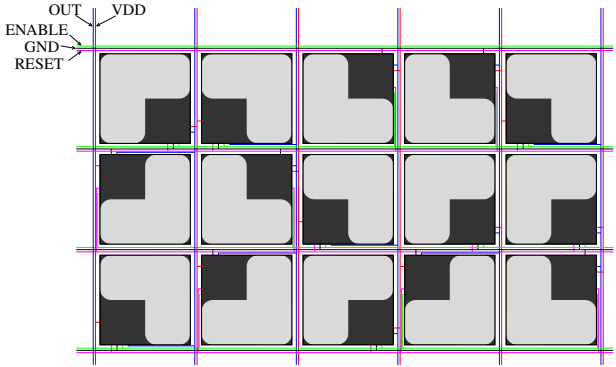}
	\caption{Small area of an image sensor with non-regularly rotated prototype pixel cells. The light-sensitive photo diodes are shown in light-gray while the insensitive circuitry is shown in dark-gray.}
	\label{fig:sensor_layout}
\end{figure}

In~\cite{Schoeberl2011}, we already have proposed a concept of modifying a regular image sensor to perform a non-regular sampling. For this, a mask is overlayed on the sensor, non-regularly shielding three quarters of every pixel. Using a fitting reconstruction algorithm we have been able to show that a higher image quality can be obtained as would be possible with the underlying regular sensor. However, this technique has the disadvantage that by shielding three quarters of every pixel, sensitivity is lost and in some cases the reconstruction can yield small artifacts.

In order to cope with both the problems, we propose a novel sensor layout which is able to capture more of the incident light and also at the same time allows for a higher quality during the reconstruction step. Looking at a typical pixel cell as shown in Figure~\ref{fig:pixel_cell}, it can be observed that the light-sensitive area forms an L-shaped region roughly covering three quarters of the pixel while the insensitive part which contains the transistors covers only one quarter. This structure of a pixel cell can be used directly for performing a non-regular sampling. While it is common up to now to just repeat this pixel cell over the whole sensor, we propose to rotate the pixel cells non-regularly as shown for a small area of the whole sensor in Figure~\ref{fig:sensor_layout}. Using this, one obtains an integration of the incident light over non-regularly rotated regions. 

Integrating the light over these non-regularly rotated \mbox{L-shaped} areas has the big advantage that it is possible to recover an image with a higher resolution by using an appropriate reconstruction algorithm. For this, two tasks have to be solved, first the integration over the spatially varying pixels has to be made undone. Second, at the position of the light-insensitive pixel area, no information is available. Thus, the signal has to be extrapolated into these unavailable regions. The sparsity-based JSDE algorithm which we propose in the next section is able to solve both problems at the same time and allows for a high reconstruction quality.

Even though the proposed sensor layout seems to fit best with pixel cells that contain an L-shaped light-sensitive area, it can also be applied to pixel cells which have a different layout or for pixel cells with micro-lenses or backside-illumination. In this case, a mask would have to be imprinted by using micro-lithography. The mask has to make one quarter of every pixel insensitive to light in a non-regular fashion, therewith achieving a layout as shown in Figure~\ref{fig:sensor_layout}.

Both possibilities, that is to say, either rotating the pixel cells or using masks are easy ways to achieve a non-regular sampling and could be included in state-of-the-art image sensor design and manufacturing processes. It is only necessary to define four prototype pixel cells with corresponding connections and place them on the image sensor plane. The only disadvantage of the proposed sensor layout is that the fill-factor of the pixel cells is limited to $75 \%$. However, as a significantly higher image quality can be achieved by the subsequent reconstruction algorithm, this is a reasonable price to pay. And for sensors which already do not make use of all the incident light, non-regularly placing the sensitive areas allows for a higher imaging quality, for free.

As mentioned above, for making use of the benefits from a non-regular sampling, an appropriate reconstruction algorithm is required. In the next section we outline the novel JSDE algorithm in detail. This algorithm is able to solve both challenges outlined above and therewith allows for the reconstruction of high-resolution images.


\section{Reconstruction by Joint Sparse Deconvolution and Extrapolation} \label{sec:reconstruction}
\begin{figure}
	\centering
	\psfrag{xt}[c][c][0.75]{$\widetilde{x}$}
	\psfrag{yt}[c][c][0.75]{$\widetilde{y}$}
	\psfrag{x}[c][c][0.75]{$x$}
	\psfrag{y}[c][c][0.75]{$y$}
	\psfrag{Reconstruction}[l][l][0.8]{Reconstruction}
	\includegraphics[width=0.48\textwidth]{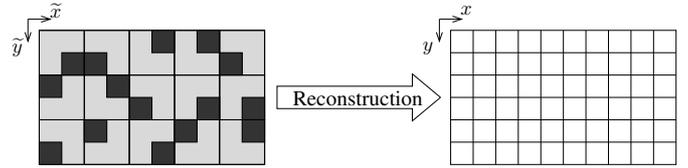}
	\caption{Fundamental idea of the reconstruction from a non-regular sampling sensor. Left: Non-regular sampling sensor (light-sensitive areas marked light-gray, insensitive areas marked dark-gray). Right: Target grid of the image to be reconstructed.}
	\label{fig:sensor_math}
\end{figure}

As discussed above, the non-regular sampling can only lead to an improved imaging quality, if it is combined with a fitting reconstruction algorithm. Therefore, we propose the sparsity-based Joint Sparse Deconvolution and Extrapolation (JSDE) in the following. For this, we will first have a look at the mathematical description of the acquisition process and the relationship between the sensor signal and the high-resolution image signal to be reconstructed. Afterwards the actual reconstruction by JSDE is outlined. 

\subsection{Mathematical Description of the Acquisition Process}

In the left half of Figure~\ref{fig:sensor_math}, a part consisting of $5\times 3$ pixels of the non-regularly sampling image sensor is shown. In general the sensor is of size $\widetilde{X}\times \widetilde{Y}$ and the output of the sensor is depicted by signal $\widetilde{s}\left[\widetilde{x}, \widetilde{y}\right]$ with spatial coordinates $\widetilde{x}$ and $\widetilde{y}$. In this context, the tilde denotes the signal with the low resolution. The objective of JSDE is to reconstruct the image on a high-resolution grid with twice the spatial resolution in vertical and horizontal direction. That is to say, every pixel of the signal $\widetilde{s}\left[\widetilde{x}, \widetilde{y}\right]$ should be expanded into four new pixels. The target signal is depicted by $s\left[x, y\right]$ and is of size $X\times Y$ with $X = 2\widetilde{X}$ and $Y = 2\widetilde{Y}$. The right side of Figure~\ref{fig:sensor_math} shows a small part consisting of $10\times 6$ pixels of the target signal to be reconstructed.

The relationship between the sensor signal $\widetilde{s}\left[\widetilde{x}, \widetilde{y}\right]$ and the target signal $s\left[x, y\right]$ to be reconstructed can be described by
\begin{eqnarray}
\widetilde{s}\left[\widetilde{x}, \widetilde{y}\right] &=& s\left[2x, 2y\right] \cdot m\left[2x, 2y\right] + \nonumber\\&& + s\left[2x+1, 2y\right] \cdot m\left[2x+1, 2y\right] +\nonumber\\&& + s\left[2x, 2y+1\right] \cdot m\left[2x, 2y+1\right] + \nonumber\\&&+ s\left[2x+1, 2y+1\right] \cdot m\left[2x+1, 2y+1\right].
\end{eqnarray}
As only three quarters of the area of every underlying pixel are sensitive to light, the output of the pixel cell results from averaging over the incident light. As the underlying pixel can be divided into four quadrants, $m\left[x, y\right]$ is $1/3$ if the underlying quadrant of the regarded pixel is sensitive to light, otherwise it is $0$. The problem to be solved for the reconstruction is recovering the signal $s\left[x, y\right]$ from $\widetilde{s}\left[\widetilde{x}, \widetilde{y}\right]$. Apparently, this is an underdetermined problem which cannot be solved directly. However, the problem can be solved by exploiting certain signal properties. For this, the proposed JSDE algorithm uses the sparsity property of image signals as shown in the following.

\subsection{Solution of the Underdetermined Problem by JSDE}

For reconstructing the signal on the high-resolution grid, JSDE first performs a division of the signal $s\left[x,y\right]$ into blocks. The signal in the block is depicted by $f\left[m,n\right]$ with spatial coordinates $m$ and $n$. A block located at position $\left(x_o, y_o\right)$ in the image signal $s\left[x,y\right]$ can be accessed by the relationship
\begin{equation}
f\left[m,n\right] = s\left[x_o+m, y_o+n\right] .
\end{equation}
The block of size $B$ together with a neighborhood of width $W$ pixels forms the reconstruction area $\mathcal{L}$. Except for cases where the considered block is located at the boundary of the image, area $\mathcal{L}$ is square and in general it is of size $M\times N$ pixels. All these coordinates and sizes are defined with respect to the fine grid. The corresponding block of the sensor signal $\widetilde{s}\left[\widetilde{x}, \widetilde{y}\right]$ is depicted by $\widetilde{f}\left[\widetilde{m},\widetilde{n}\right]$ and is half the size in vertical and horizontal direction compared to the block $f\left[m,n\right]$ from the high-resolution signal.

\begin{figure}
	\centering
	\includegraphics[width=0.2\textwidth]{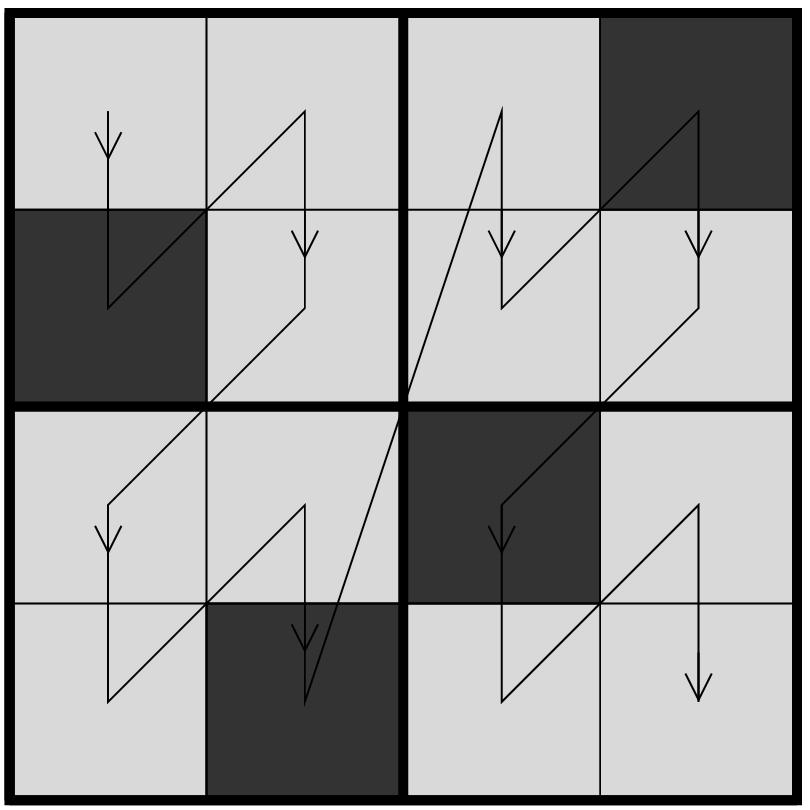}
	\caption{Scan order for four pixel for vectorizing $f\left[m,n\right]$ to $\ve{f}$. The light-insensitive areas of the underlying large pixels are indicated by dark-gray.}
	\label{fig:scan_order}
\end{figure}

In order to describe the relationship between the non-regularly sampled signal and the target high-resolution signal, the signals are vectorized. By scanning the sensor signal $\widetilde{f}\left[\widetilde{m},\widetilde{n}\right]$ in column order, we obtain vector $\widetilde{\ve{f}}$. Accordingly, all the pixels of $f\left[m,n\right]$ are aligned in the vector $\ve{f}$. However, it has to be noted that the mapping between the two-dimensional signal $f\left[m,n\right]$ and the vector $\ve{f}$ is not column-wise, but rather, first inside every of the related large pixels a scan is performed column-wise before the scan proceeds to the next pixel. Figure~\ref{fig:scan_order} illustrates this scan order.

\begin{figure}
	\centering
	\psfrag{f}[c][c][0.75]{$\ve{f}^\trans$}
	\psfrag{ft}[c][c][0.75]{$\widetilde{\ve{f}}^\trans$}
	\includegraphics[width=0.4\textwidth]{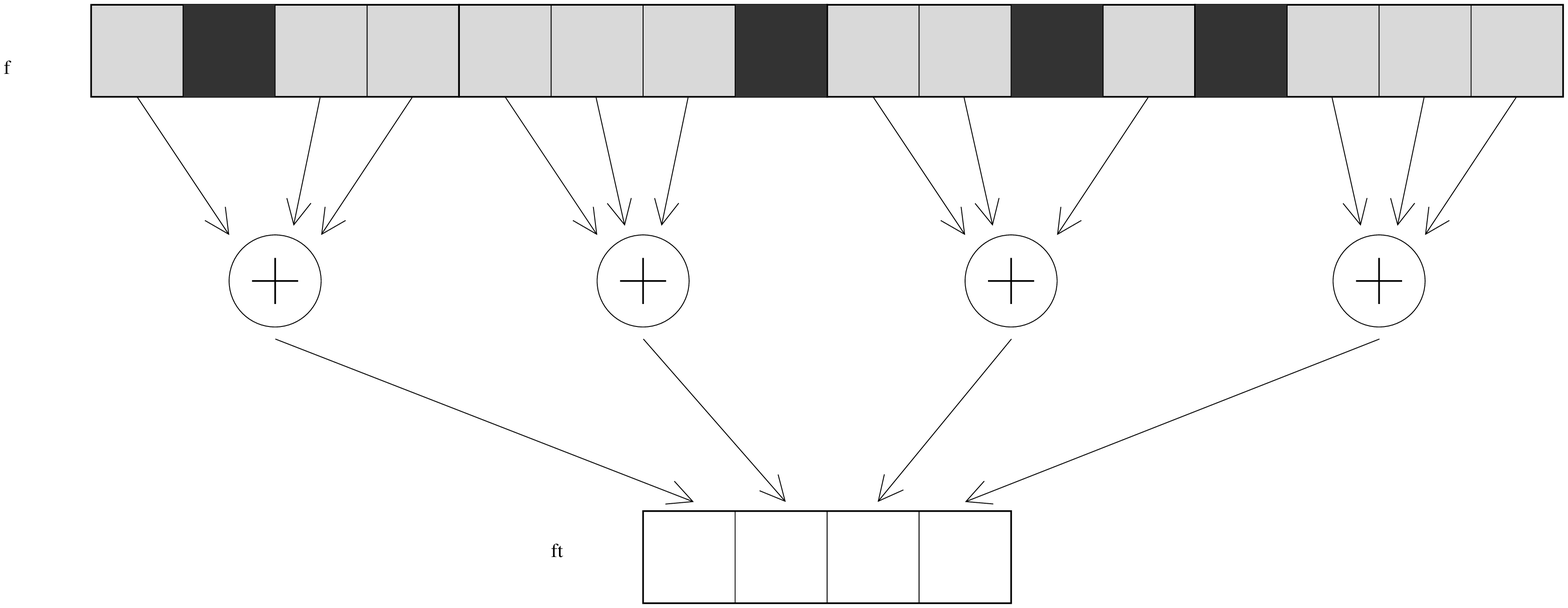}
	\caption{Relationship between sensor signal $\widetilde{\ve{f}}$ and desired high-resolution signal $\ve{f}$.}
	\label{fig:vector_aggregation}
\end{figure}

As every pixel of the desired high-resolution image relates to one quadrant of one pixel of the low-resolution sensor signal, every time three of the four pixels of the high-resolution signal contribute to one pixel of the sensor signal. The relationship between $\widetilde{\ve{f}}$ and $\ve{f}$ is shown in Figure~\ref{fig:vector_aggregation} and can also be described by the block diagonal aggregation matrix $\ve{A}$, leading to
\begin{equation}
\widetilde{\ve{f}} = \ve{A}\ve{f}.
\end{equation}
Regarding the examples shown in Figures~\ref{fig:scan_order} and~\ref{fig:vector_aggregation}, the corresponding aggregation matrix $\ve{A}$ is given by
\begin{equation}
\ve{A}\hspace{-1mm} =\hspace{-1mm} \frac{1}{3}\hspace{-1mm}\left(\hspace{-1mm} \begin{array}{cccc} 1\ \ 0\ \ 1\ \ 1 &&\ve{0}& \\ & 1\ \ 1\ \ 1\ \ 0 & &\\ && 1\ \ 1\ \ 0\ \ 1 & \\ &\ve{0}&& 0\ \ 1\ \ 1\ \ 1\end{array}\hspace{-1mm}\right)\hspace{-1mm}.
\end{equation}
Of course, for the aggregation of the whole signal $\ve{f}$ to $\widetilde{\ve{f}}$, matrix $\ve{A}$ is larger and it always is of size $\frac{MN}{4} \times \left(MN\right)$.

Apparently, it is not possible to recover $\ve{f}$ from $\widetilde{\ve{f}}$ directly, since $\ve{A}$ cannot be inverted. As stated above, the inversion is an underdetermined problem consisting of two sub-problems to be solved. First, a spatial varying deconvolution has to be performed which inverts the integration of the three light-sensitive quadrants into the output of the pixel of the image sensor. And second, the signal amplitude at the position where the sensor is insensitive to light has to be estimated which can be regarded as extrapolation problem. For solving both tasks and for recovering the signal on the high-resolution grid, we propose the novel JSDE which exploits the fact that image signals can be sparsely represented in the frequency domain~\cite{Candes2007}.

For the reconstruction, JSDE aims at generating the parametric model
\begin{equation}
g\left[m,n\right] = \sum_{k\in \mathcal{K}} c_k \varphi_k\left[m,n\right]
\end{equation}
as a superposition of two-dimensional Fourier basis functions $\varphi_k\left[m,n\right]$. The weights of the individual basis functions are depicted by the expansion coefficients $c_k$ and the indices of all basis functions used for the model generation are subsumed in set $\mathcal{K}$. The Fourier basis functions are defined by
\begin{equation}
\label{eq:fourier_bf}
\varphi_{k}\left[m,n\right] = \e^{\frac{2\pi\j}{M}\mathrm{mod}\left(k,M\right)m} \e^{\frac{2\pi\j}{N}\lfloor k/M \rfloor n}.
\end{equation}
By using the same scan order as shown in Figure~\ref{fig:scan_order}, the model can also be vectorized to $\ve{g}$ and the basis functions to $\ve{\varphi}_k$.

For generating the model, JSDE uses an iterative approach that is related to the Frequency Selective Reconstruction (FSR)~\cite{Seiler2015} and its precursor Frequency Selective Extrapolation~\cite{Seiler2010c}. However, it has to be noted that FSR only is able to perform an extrapolation while JSDE can additionally perform a spatial varying deconvolution and therewith can be regarded as a generalization of FSR. In order to generate the model, the distribution matrix $\ve{D}$ is required that controls which pixels of signal $\ve{f}$ belong to pixels from the low-resolution signal $\widetilde{\ve{f}}$. Distribution matrix $\ve{D}$ is closely related to aggregation matrix $\ve{A}$ and also is a block diagonal matrix. However, $\ve{D}$ carries blocks of four ones and can be used for directly expanding $\widetilde{\ve{f}}$ to the high-resolution grid and therewith can be regarded as the dual operation to aggregation matrix $\ve{A}$. Hence, $\ve{D}$ is of size $\left(MN\right) \times \frac{MN}{4}$. For the example shown in Figures~\ref{fig:scan_order} and~\ref{fig:vector_aggregation}, distribution matrix $\ve{D}$ looks as follows:
\begin{equation}
\ve{D} = \left( \begin{array}{cccc} 1 &&&\\ 1 &&&\\ 1 &&&\\ 1 &&\ve{0}&\\ & 1 &&\\ & 1 &&\\ & 1 &&\\ & 1 &&\\ && 1 &\\ && 1 &\\ && 1 &\\ && 1 &\\ &\ve{0}&& 1\\ &&& 1\\ &&& 1\\ &&& 1\end{array}\right)
\end{equation}
Using the aggregation matrix $\ve{A}$ and the distribution matrix $\ve{D}$, the residual $\ve{r}$ between the available signal $\widetilde{\ve{f}}$ and the model $\ve{g}$ can be determined with respect to the high-resolution grid.
\begin{equation}
\ve{r} = \ve{D} \widetilde{\ve{f}} - \ve{D}\ve{A}\ve{g}
\end{equation}

During the model generation process, the spatial weighting function 
\begin{equation}
\label{eq:weighting_function}
w\left[m,n\right] = \left\{ \hspace{-1mm}\begin{array}{ll} \hat{\rho}^{\sqrt{\left(m-\frac{M-1}{2}\right)^2+\left(n-\frac{N-1}{2}\right)^2}} & \mbox{for } \left(m,n\right)\in \mathcal{A} \\ 0 & \mbox{for } \left(m,n\right)\in \mathcal{B}\end{array}\right.
\end{equation}
similar to~\cite{Meisinger2004b} is used. Here, set $\mathcal{A}$ subsumes all the pixels relating to light-sensitive quadrants, whereas set $\mathcal{B}$ contains the insensitive quadrants. This weighting function is required for two reasons. First, the pixels from the fine grid which relate to the light-insensitive areas can be excluded from the model generation. Since these pixels do not contribute to the measured amplitudes in the underlying low-resolution pixels, they contain no information and have to be excluded from the model generation. Second, a weighting with an isotropic model is carried out in order to assign a lower weight and therewith less influence on the model generation to pixels lying more far away from the block to be reconstructed than pixels located closer to the considered block. 

For generating the model of the signal, the weighted residual energy $E_w$ is considered. $E_w$ describes how well the generated model on the fine grid fits the available signal on the coarse grid, subject to the weighting function $w\left[m,n\right]$. By scanning $w\left[m,n\right]$ in the same way as $f\left[m,n\right]$ shown in Figure~\ref{fig:scan_order}, we obtain vector $\ve{w}$ and the diagonal matrix $\ve{W} = \diag\left(\ve{w}\right)$. Using this, the weighted residual energy can be calculated by 
\begin{eqnarray}
\nonumber E_w &=& \ve{r}^\herm\ve{W}\ve{r} \\
&=& \left(\ve{D} \widetilde{\ve{f}} - \ve{D}\ve{A}\ve{g}\right)^\herm \ve{W} \left(\ve{D} \widetilde{\ve{f}} - \ve{D}\ve{A}\ve{g}\right).
\end{eqnarray}
In this context, $\left(\cdot\right)^\herm$ denotes the conjugate transpose, or respectively the Hermitian matrix or vector.

As mentioned above, the model $\ve{g}$ will be generated iteratively and the currently considered iteration is depicted by $\nu$. Initially, the model is set to zero, $\ve{g}^{\left(0\right)} = \ve{0}$, and the residual is set to the expanded low-resolution signal $\ve{r}^{\left(0\right)} = \ve{D}\widetilde{\ve{f}}$. Then, in every iteration one basis function to be added to the model is selected and the corresponding expansion coefficient is estimated. In order to determine the basis function to be added to the model in the current iteration, a weighted projection of the residual onto the basis function is carried out. For calculating the weighted projection, the weighted residual energy $E_{w,k}^{\left(\nu\right)}$ is regarded, which would result from a selection of basis function $\ve{\varphi}_k$ in iteration $\nu$. The weighted residual energy can be calculated by 
\begin{equation}
E_{w,k}^{\left(\nu\right)} = \ve{r}_k^{\left(\nu\right)^\herm} \ve{W} \ve{r}_k^{\left(\nu\right)}
\end{equation}
with the residual 
\begin{equation}
\ve{r}_k^{\left(\nu\right)} =\ve{r}^{\left(\nu-1\right)} - p_k^{\left(\nu\right)} \ve{D}\ve{A}\ve{\varphi}_k
\end{equation}
which would result if basis function $\ve{\varphi}_k$ was selected in iteration $\nu$. Based on this, the projection coefficients $p_k^{\left(\nu\right)}$ as output from the weighted projection can be calculated for all indices $k$.

The projection coefficient $p_k^{\left(\nu\right)}$ is determined by minimizing $E_{w,k}^{\left(\nu\right)}$. For this, the Wirtinger calculus is used and the partial derivatives
\begin{equation}
\label{eq:Wirtinger_calculus}
\frac{\partial E_{w,k}^{\left(\nu\right)}}{\partial p_k^{\left(\nu\right)}} \stackrel{!}{=} 0 \hspace{5mm}\mbox{and}\hspace{5mm} \frac{\partial E_{w,k}^{\left(\nu\right)}}{\partial p_k^{\left(\nu\right)^\ast}} \stackrel{!}{=} 0
\end{equation}
are set to zero. Here, the $\left(\cdot\right)^\ast$ denotes the conjugate complex term. The partial derivative $\frac{\partial E_{w,k}^{\left(\nu\right)}}{\partial p_k^{\left(\nu\right)^\ast}}$ leads to
\[
\frac{\partial E_{w,k}^{\left(\nu\right)}}{\partial p_k^{\left(\nu\right)^\ast}} = \frac{\partial }{\partial p_k^{\left(\nu\right)^\ast}} \left(\ve{r}_k^{\left(\nu\right)^\herm} \ve{W} \ve{r}_k^{\left(\nu\right)}\right) \hspace{4cm}
\]
\begin{equation}
\hspace{5mm}=\hspace{-1mm} \frac{1}{2}\hspace{-1mm} \left(\hspace{-1mm} -\left(\ve{D}\ve{A}\ve{\varphi}_k\right)^\herm\ve{W}\ve{r}^{\left(\nu-1\right)} \hspace{-1mm}+\hspace{-1mm} p_k^{\left(\nu\right)} \left(\ve{D}\ve{A}\ve{\varphi}_k\right)^\herm\ve{W}\ve{D}\ve{A}\ve{\varphi}_k\hspace{-1mm} \right) .
\end{equation}
In order to fulfill (\ref{eq:Wirtinger_calculus}) for minimizing the weighted residual energy,
\begin{equation}
p_k^{\left(\nu\right)} = \frac{\left(\ve{D}\ve{A}\ve{\varphi}_k\right)^\herm\ve{W}\ve{r}^{\left(\nu-1\right)}}{\left(\ve{D}\ve{A}\ve{\varphi}_k\right)^\herm\ve{W}\ve{D}\ve{A}\ve{\varphi}_k}
\end{equation}
follows as result for the projection coefficient.

\begin{figure}
	\centering
	
\psfrag{s02}[rt][rt]{\color[rgb]{0,0,0}\setlength{\tabcolsep}{0pt}\begin{tabular}{r}$\widetilde{k_1}$\end{tabular}}%
\psfrag{s03}[lt][lt]{\color[rgb]{0,0,0}\setlength{\tabcolsep}{0pt}\begin{tabular}{l}$\widetilde{k_2}$\end{tabular}}%
\psfrag{s04}[b][b]{\color[rgb]{0,0,0}\setlength{\tabcolsep}{0pt}\begin{tabular}{c}$q_k$\end{tabular}}%

\psfrag{x01}[t][t][0.75]{$-15$}%
\psfrag{x02}[t][t][0.75]{$-10$}%
\psfrag{x03}[t][t][0.75]{$-5$}%
\psfrag{x04}[t][t][0.75]{$0$}%
\psfrag{x05}[t][t][0.75]{$5$}%
\psfrag{x06}[t][t][0.75]{$10$}%
\psfrag{x07}[t][t][0.75]{$15$}%
\psfrag{v01}[r][r][0.75]{$-15$}%
\psfrag{v02}[r][r][0.75]{$-10$}%
\psfrag{v03}[r][r][0.75]{$-5$}%
\psfrag{v04}[r][r][0.75]{$0$}%
\psfrag{v05}[r][r][0.75]{$5$}%
\psfrag{v06}[r][r][0.75]{$10$}%
\psfrag{v07}[r][r][0.75]{$15$}%
\psfrag{z01}[r][r][0.75]{$0$}%
\psfrag{z02}[r][r][0.75]{$0.2$}%
\psfrag{z03}[r][r][0.75]{$0.4$}%
\psfrag{z04}[r][r][0.75]{$0.6$}%
\psfrag{z05}[r][r][0.75]{$0.8$}%
\psfrag{z06}[r][r][0.75]{$1$}%
	
	\includegraphics[width=0.4\textwidth]{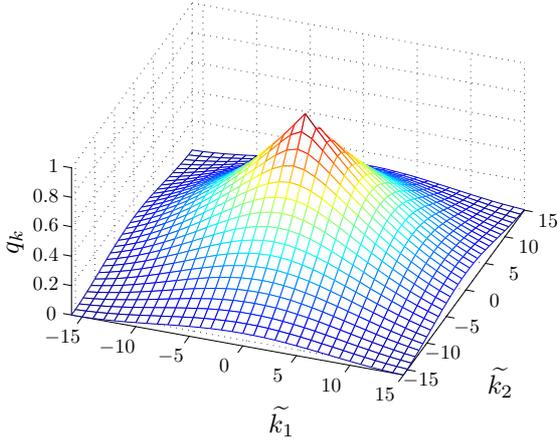}
	\caption{Two-dimensional plot of frequency prior $q_k$ with respect to substitute variables $\widetilde{k_1}$ and $\widetilde{k_2}$.}
	\label{fig:frequency_prior}
\end{figure}

For determining which basis function actually to add in iteration $\nu$, the one is selected that would be able to reduce the weighted residual energy the most, subject to the frequency prior $q_k$. The frequency prior is used for favoring the selection of low-frequency basis functions over high-frequency basis functions in the case of ambiguities. The influence of the frequency prior on the model generation is explained in detail in~\cite{Seiler2015, Seiler2016}. While in~\cite{Seiler2015} a prior inspired by the optical transfer function (OTF) of imaging systems is used, an adaptive frequency prior is proposed in~\cite{Seiler2016}. For JSDE, the simple OTF-inspired prior is sufficient and $q_k$ is defined by
\begin{equation}
q_k = \left( 1 - \sqrt{2} \sqrt{\frac{\widetilde{k_1}^2}{M^2} + \frac{\widetilde{k_2}^2}{N^2}} \right)^2.
\end{equation} 
Here, the two substitutions $\widetilde{k_1} = \frac{M}{2} - \left|\mathrm{mod}\left(k,M\right) - \frac{M}{2}\right|$ and $\widetilde{k_2} = \frac{N}{2} - \left|\lfloor k/N \rfloor-\frac{N}{2}\right|$ are used for allowing a compact representation. A two-dimensional plot of the frequency prior $q_k$ with respect to the substitute variables $\widetilde{k_1}$ and $\widetilde{k_2}$ is provided in Figure~\ref{fig:frequency_prior}.

Using this, the basis function to select is the one that minimizes $E_{w,k}^{\left(\nu\right)}$ subject to the frequency prior and the index $u^{\left(\nu\right)}$ of the basis function results to
\begin{eqnarray}
\label{eq:bf_selection_origin} u^{\left(\nu\right)} &=& \argmin_k \left(q_k E_{w,k}^{\left(\nu\right)}\right) \\
\label{eq:bf_selection}&=& \argmax_k \left(q_k \frac{\left|\left(\ve{D}\ve{A}\ve{\varphi}_k\right)^\herm\ve{W}\ve{r}^{\left(\nu-1\right)}\right|^2}{\left(\ve{D}\ve{A}\ve{\varphi}_k\right)^\herm\ve{W}\ve{D}\ve{A}\ve{\varphi}_k}\right).
\end{eqnarray}
The steps for obtaining the result above can be found in the Appendix. 

After having selected the basis function to be added in the current iteration, the update of the expansion coefficient has to be determined. For this, the orthogonality deficiency compensation proposed in~\cite{Seiler2007,Seiler2008} is used. This procedure is used for reducing the interference between the different basis functions and therewith obtaining a stable estimation. For this, only a portion of the projection coefficient is added to the expansion coefficient. Even though directly using the projection coefficient for the update would reduce the residual energy the most, reducing the influence by orthogonality deficiency compensation yields a better modeling. As shown in~\cite{Seiler2008}, a constant factor $\gamma$ could be used as a good approximation for the elaborate compensation of the orthogonality deficiency which is proposed in~\cite{Seiler2007}. Using this, the model and the residual can be updated by
\begin{eqnarray}
\ve{g}^{\left(\nu\right)} &=& \ve{g}^{\left(\nu-1\right)} + \gamma p_{u^{\left(\nu\right)}}^{\left(\nu\right)}\ve{\varphi}_{u^{\left(\nu\right)}} \\
\ve{r}^{\left(\nu\right)} &=& \ve{r}^{\left(\nu-1\right)} - \gamma p_{u^{\left(\nu\right)}}^{\left(\nu\right)}\ve{D}\ve{A}\ve{\varphi}_{u^{\left(\nu\right)}}.
\end{eqnarray}

The steps of selecting the basis function to be added, estimating the weight and updating the model and residual are repeated for a pre-defined number of iterations $I$. After the model generation has been finished, the samples corresponding to the currently considered block are extracted from the model and are placed in the high-resolution image. Finally the reconstruction proceeds to the next block. In order to provide a compact overview of the model generation of JSDE, Algorithm~\ref{algo:model_generation} shows a pseudo code of the modeling.

\begin{algorithm}[t]
	\caption{\emph{Pseudo code for model generation of the Joint Sparse Deconvolution and Extrapolation algorithm. Signals are vectorized according to the scan order shown in Figure~\ref{fig:scan_order}.}}
	\label{algo:model_generation}
	\begin{algorithmic}
		\INPUT Block $\widetilde{\ve{f}}$ of low-resolution signal, sizes $M$ and $N$, aggregation matrix $\ve{A}$, distribution matrix $\ve{D}$, and weighting matrix $\ve{W}$ according to sensor layout, basis functions $\ve{\varphi}_k$, number of iterations $I$, and compensation factor $\gamma$.
		\STATE /* Initialization */
		\STATE $\ve{g}^{\left(0\right)} = \ve{0}$ 
		\STATE $\ve{r}^{\left(0\right)} = \ve{D}\widetilde{\ve{f}}$
		\FORALL {$k=0, \ldots, MN-1$}
		\STATE $\widetilde{k_1} = \frac{M}{2} - \left|\mathrm{mod}\left(k,M\right) - \frac{M}{2}\right|$
		\STATE $\widetilde{k_2} = \frac{N}{2} - \left|\lfloor k/N \rfloor-\frac{N}{2}\right|$
		\STATE $q_k = \left( 1 - \sqrt{2} \sqrt{\frac{\widetilde{k_1}^2}{M^2} + \frac{\widetilde{k_2}^2}{N^2}} \right)^2$
		\ENDFOR	
		\STATE /* Iterative model generation */
		\FORALL {$\nu=1, \ldots, I$}		
		\STATE /* Basis function selection */
		\STATE $u^{\left(\nu\right)} = \argmax_k \left(q_k \frac{\left|\left(\ve{D}\ve{A}\ve{\varphi}_k\right)^\herm\ve{W}\ve{r}^{\left(\nu-1\right)}\right|^2}{\left(\ve{D}\ve{A}\ve{\varphi}_k\right)^\herm\ve{W}\ve{D}\ve{A}\ve{\varphi}_k}\right)$
		\STATE /* Projection coefficient calculation */
		\STATE $p_{u^{\left(\nu\right)}}^{\left(\nu\right)} = \frac{\left(\ve{D}\ve{A}\ve{\varphi}_{u^{\left(\nu\right)}}\right)^\herm\ve{W}\ve{r}^{\left(\nu-1\right)}}{\left(\ve{D}\ve{A}\ve{\varphi}_{u^{\left(\nu\right)}}\right)^\herm\ve{W}\ve{D}\ve{A}\ve{\varphi}_{u^{\left(\nu\right)}}}$
		\STATE /* Model and residual update */
		\STATE $\ve{g}^{\left(\nu\right)} = \ve{g}^{\left(\nu-1\right)} + \gamma p_{u^{\left(\nu\right)}}^{\left(\nu\right)}\ve{\varphi}_{u^{\left(\nu\right)}}$
		\STATE $\ve{r}^{\left(\nu\right)} = \ve{r}^{\left(\nu-1\right)} - \gamma p_{u^{\left(\nu\right)}}^{\left(\nu\right)}\ve{D}\ve{A}\ve{\varphi}_{u^{\left(\nu\right)}}$
		\ENDFOR
		\OUTPUT Model $\ve{g} = \ve{g}^{\left(I\right)}$ with respect to the fine grid
		
	\end{algorithmic}
\end{algorithm}

\subsection{Relationship to Compressed Sensing}

Looking at the model generation process of JSDE outlined above, it can be observed that there are several similarities to reconstruction algorithms used within the Compressed Sensing (CS) framework~\cite{Candes2006, Donoho2006}. Indeed, JSDE can also be seen in this framework as it generates a sparse model of the signal that is measured. In this context, the aggregation matrix $\ve{A}$ could be interpreted as the sensing matrix and the pixel amplitudes of the low-resolution sensors can be regarded as the output from the linear measurements of the high-resolution signal. 

In~\cite{Seiler2015}, the relationship between the FSR algorithm and CS is discussed in detail. As FSR can be regarded as a precursor of the proposed JSDE, this relationship also holds for JSDE. Accordingly, JSDE belongs to the group of greedy algorithms whose most prominent representative is Matching Pursuits (MP)~\cite{Mallat1993}. However, by incorporating the spatial weighting function and the frequency prior, JSDE is also related to CS algorithms that make use of prior knowledge, as for example the ones proposed in~\cite{Masood2013, Chi2013}. Furthermore, the use of a block-wise processing makes JSDE also related to block-wise CS algorithms~\cite{Mun2009}, especially the ones that use overlapping blocks~\cite{Gutierrez2014}.

In the next section, simulation results are provided for showing the effectiveness of JSDE and its ability to recover a high-resolution image from a non-regular sampling sensor. We also show that the combination of non-regular sampling and a fitting reconstruction yields a superior quality and is able to outperform different sampling and reconstruction concepts.


\section{Simulations and Results}\label{sec:simulations} 
\begin{figure}
	\centering
	\psfrag{Large pixel}[l][l][0.75]{Large pixel}
	\psfrag{Non-regular quarter sampling}[l][l][0.75]{Non-regular quarter sampling}
	\psfrag{Regular three-quarter sampling}[l][l][0.75]{Regular three-quarter sampling}
	\psfrag{Non-regular three-quarter sampling}[l][l][0.75]{Non-regular three-quarter sampling}
	\includegraphics[width=0.45\textwidth]{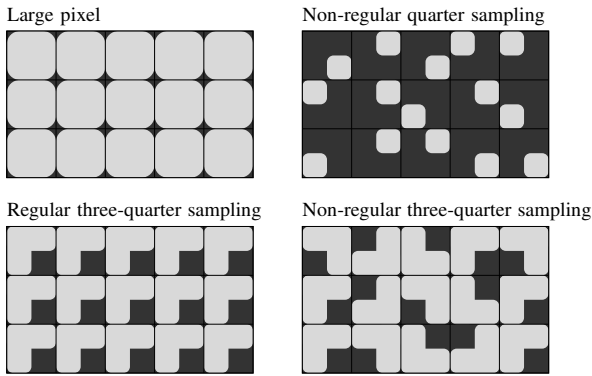}
	\caption{Considered sensor layouts: Large pixels with $\approx 100\%$ fill-factor, non-regular quarter sampling~\cite{Schoeberl2011} with $\approx 25\%$ fill-factor, regular three-quarter sampling with $\approx 75\%$ fill-factor, and the proposed non-regular three-quarter sampling with $\approx 75\%$ fill-factor.}
	\label{fig:test_sensor_layouts}
\end{figure}

\subsection{Simulation Setup}

The purpose of this section is to show how well the proposed non-regular sampling in combination with the reconstruction by JSDE can be used for achieving a higher imaging quality. In order to obtain a meaningful evaluation, this section also provides simulation results for different sensor layouts in combination with various reconstruction algorithms. As layouts for the sensor, four different schemes are considered, as shown in Figure~\ref{fig:test_sensor_layouts}. First, these are large pixels with $\approx 100 \%$ fill-factor which could be achieved for example if backside-illumination or micro lenses are used. Second, the non-regular quarter sampling~\cite{Schoeberl2011} with $25 \%$ fill-factor is considered. Furthermore, two different cases are examined, where a three-quarter sampling with $\approx 75 \%$ fill-factor is considered. This relates to the use of a prototype pixel cell as shown in Figure~\ref{fig:pixel_cell} without microlenses. This protoype pixel cell can either be identically repeated or, as proposed, non-regularly rotated in order to place the insensitive area in different quadrants.

In order to simulate the behavior of the different sensing technologies, high-resolution images are taken and every time $2\times 2$ pixels of this image are combined. Depending on the considered sensor layout, the four pixels are either just averaged as for the large pixel case, or one of the four pixels is selected for the non-regular quarter sampling~\cite{Schoeberl2011} case. For the latter two cases, three of the four samples are averaged while the fourth is discarded. For the regular sampling case, always the same three pixels of a  $2\times 2$ group are averaged while for the non-regular sampling case, the considered pixels over which the averaging is carried out change.

In order to reconstruct the high-resolution images from the sensor output, different algorithms are evaluated. Of course, which algorithm to actually use depends on the regarded sensor layout. For the large pixels case, where the light-sensitive areas are square and extend over the whole pixels, the easiest way to obtain an image on the high-resolution grid just would be to apply a pixel enlargement (PE). That is to say, the amplitude of the measured pixel is assigned to all the four underlying pixels on the high-resolution grid. This can also be regarded as a nearest neighbor interpolation. Alternative to this, we also test a bicubic upsampling (BIC) by a factor of two in both directions. As the objective is to reconstruct the image on a grid of a higher resolution and this task can also be solved by SR algorithms, we also include three single image SR algorithms into the evaluation. These are the sparse dictionary-based algorithms from Yang (SR-Yang) \cite{Yang2010} and Zeyde (SR-Zeyde)~\cite{Zeyde2010} on the one hand side and the regression-based algorithm from Kim (SR-Kim)~\cite{Kim2010a} on the other hand side.

For the case that the sensor layout using the non-regular quarter sampling~\cite{Schoeberl2011} is used, an extrapolation is required for reconstructing the missing pixels with respect to the high-resolution grid. For this FSR~\cite{Seiler2015} is considered as it can be seen as an ancestor of JSDE. Furthermore, Steering Kernel Regression (KR)~\cite{Takeda2007} and Constrained Split Augmented Lagrangian Shrinkage Algorithm (CLS)~\cite{Afonso2011} are applied for solving this task.

\begin{figure*} 
	\centering
	\begin{minipage}{0.32\textwidth}



\providelength{\AxesLineWidth}       \setlength{\AxesLineWidth}{0.5pt}
\providelength{\GridLineWidth}       \setlength{\GridLineWidth}{0.4pt}
\providelength{\GridLineDotSep}      \setlength{\GridLineDotSep}{0.4pt}
\providelength{\MinorGridLineWidth}  \setlength{\MinorGridLineWidth}{0.4pt}
\providelength{\MinorGridLineDotSep} \setlength{\MinorGridLineDotSep}{0.8pt}
\providelength{\plotwidth}           \setlength{\plotwidth}{4.5cm} 
\providelength{\LineWidth}           \setlength{\LineWidth}{0.7pt}
\providelength{\MarkerSize}          \setlength{\MarkerSize}{4pt}
\newrgbcolor{GridColor}{0.8 0.8 0.8}

\psset{xunit=0.004000\plotwidth,yunit=0.266667\plotwidth}
\begin{pspicture}(-41.666667,28.375000)(250.000000,32.000000)


\psline[linestyle=dashed,dash=2pt 3pt,dotsep=\GridLineDotSep,linewidth=\GridLineWidth,linecolor=GridColor](0.000000,29.000000)(0.000000,32.000000)
\psline[linestyle=dashed,dash=2pt 3pt,dotsep=\GridLineDotSep,linewidth=\GridLineWidth,linecolor=GridColor](50.000000,29.000000)(50.000000,32.000000)
\psline[linestyle=dashed,dash=2pt 3pt,dotsep=\GridLineDotSep,linewidth=\GridLineWidth,linecolor=GridColor](100.000000,29.000000)(100.000000,32.000000)
\psline[linestyle=dashed,dash=2pt 3pt,dotsep=\GridLineDotSep,linewidth=\GridLineWidth,linecolor=GridColor](150.000000,29.000000)(150.000000,32.000000)
\psline[linestyle=dashed,dash=2pt 3pt,dotsep=\GridLineDotSep,linewidth=\GridLineWidth,linecolor=GridColor](200.000000,29.000000)(200.000000,32.000000)
\psline[linestyle=dashed,dash=2pt 3pt,dotsep=\GridLineDotSep,linewidth=\GridLineWidth,linecolor=GridColor](250.000000,29.000000)(250.000000,32.000000)
\psline[linestyle=dashed,dash=2pt 3pt,dotsep=\GridLineDotSep,linewidth=\GridLineWidth,linecolor=GridColor](0.000000,29.000000)(250.000000,29.000000)
\psline[linestyle=dashed,dash=2pt 3pt,dotsep=\GridLineDotSep,linewidth=\GridLineWidth,linecolor=GridColor](0.000000,29.500000)(250.000000,29.500000)
\psline[linestyle=dashed,dash=2pt 3pt,dotsep=\GridLineDotSep,linewidth=\GridLineWidth,linecolor=GridColor](0.000000,30.000000)(250.000000,30.000000)
\psline[linestyle=dashed,dash=2pt 3pt,dotsep=\GridLineDotSep,linewidth=\GridLineWidth,linecolor=GridColor](0.000000,30.500000)(250.000000,30.500000)
\psline[linestyle=dashed,dash=2pt 3pt,dotsep=\GridLineDotSep,linewidth=\GridLineWidth,linecolor=GridColor](0.000000,31.000000)(250.000000,31.000000)
\psline[linestyle=dashed,dash=2pt 3pt,dotsep=\GridLineDotSep,linewidth=\GridLineWidth,linecolor=GridColor](0.000000,31.500000)(250.000000,31.500000)
\psline[linestyle=dashed,dash=2pt 3pt,dotsep=\GridLineDotSep,linewidth=\GridLineWidth,linecolor=GridColor](0.000000,32.000000)(250.000000,32.000000)

\psline[linewidth=\AxesLineWidth,linecolor=GridColor](0.000000,29.000000)(0.000000,29.045000)
\psline[linewidth=\AxesLineWidth,linecolor=GridColor](50.000000,29.000000)(50.000000,29.045000)
\psline[linewidth=\AxesLineWidth,linecolor=GridColor](100.000000,29.000000)(100.000000,29.045000)
\psline[linewidth=\AxesLineWidth,linecolor=GridColor](150.000000,29.000000)(150.000000,29.045000)
\psline[linewidth=\AxesLineWidth,linecolor=GridColor](200.000000,29.000000)(200.000000,29.045000)
\psline[linewidth=\AxesLineWidth,linecolor=GridColor](250.000000,29.000000)(250.000000,29.045000)
\psline[linewidth=\AxesLineWidth,linecolor=GridColor](0.000000,29.000000)(3.000000,29.000000)
\psline[linewidth=\AxesLineWidth,linecolor=GridColor](0.000000,29.500000)(3.000000,29.500000)
\psline[linewidth=\AxesLineWidth,linecolor=GridColor](0.000000,30.000000)(3.000000,30.000000)
\psline[linewidth=\AxesLineWidth,linecolor=GridColor](0.000000,30.500000)(3.000000,30.500000)
\psline[linewidth=\AxesLineWidth,linecolor=GridColor](0.000000,31.000000)(3.000000,31.000000)
\psline[linewidth=\AxesLineWidth,linecolor=GridColor](0.000000,31.500000)(3.000000,31.500000)
\psline[linewidth=\AxesLineWidth,linecolor=GridColor](0.000000,32.000000)(3.000000,32.000000)

{ \footnotesize 
\rput[t](0.000000,28.955000){$0$}
\rput[t](50.000000,28.955000){$50$}
\rput[t](100.000000,28.955000){$100$}
\rput[t](150.000000,28.955000){$150$}
\rput[t](200.000000,28.955000){$200$}
\rput[t](250.000000,28.955000){$250$}
\rput[r](-3.000000,29.000000){$29$}
\rput[r](-3.000000,30.000000){$30$}
\rput[r](-3.000000,31.000000){$31$}
\rput[r](-3.000000,32.000000){$32$}
} 

\pspolygon[linewidth=\AxesLineWidth](0.000000,29.000000)(250.000000,29.000000)(250.000000,32.000000)(0.000000,32.000000)(0.000000,29.000000)

{ \small 
\rput[b](125.000000,28.375000){
\begin{tabular}{c}
$I$\\
\end{tabular}
}

\rput[t]{90}(-41.666667,30.500000){
\begin{tabular}{c}
PSNR [dB]\\
\end{tabular}
}
} 

\newrgbcolor{color180.0083}{1  0  0}
\savedata{\mydata}[{
{25.000000,30.040000},{50.000000,31.170000},{100.000000,31.560000},{150.000000,31.600000},{250.000000,31.560000},
}]
\dataplot[plotstyle=line,linestyle=solid,linewidth=\LineWidth,linecolor=color180.0083]{\mydata}

\end{pspicture}%
	\end{minipage}
	\begin{minipage}{0.32\textwidth}



\providelength{\AxesLineWidth}       \setlength{\AxesLineWidth}{0.5pt}
\providelength{\GridLineWidth}       \setlength{\GridLineWidth}{0.4pt}
\providelength{\GridLineDotSep}      \setlength{\GridLineDotSep}{0.4pt}
\providelength{\MinorGridLineWidth}  \setlength{\MinorGridLineWidth}{0.4pt}
\providelength{\MinorGridLineDotSep} \setlength{\MinorGridLineDotSep}{0.8pt}
\providelength{\plotwidth}           \setlength{\plotwidth}{4.5cm} 
\providelength{\LineWidth}           \setlength{\LineWidth}{0.7pt}
\providelength{\MarkerSize}          \setlength{\MarkerSize}{4pt}
\newrgbcolor{GridColor}{0.8 0.8 0.8}

\psset{xunit=1.250000\plotwidth,yunit=0.266667\plotwidth}
\begin{pspicture}(0.066667,28.375000)(1.000000,32.000000)


\psline[linestyle=dashed,dash=2pt 3pt,dotsep=\GridLineDotSep,linewidth=\GridLineWidth,linecolor=GridColor](0.200000,29.000000)(0.200000,32.000000)
\psline[linestyle=dashed,dash=2pt 3pt,dotsep=\GridLineDotSep,linewidth=\GridLineWidth,linecolor=GridColor](0.300000,29.000000)(0.300000,32.000000)
\psline[linestyle=dashed,dash=2pt 3pt,dotsep=\GridLineDotSep,linewidth=\GridLineWidth,linecolor=GridColor](0.400000,29.000000)(0.400000,32.000000)
\psline[linestyle=dashed,dash=2pt 3pt,dotsep=\GridLineDotSep,linewidth=\GridLineWidth,linecolor=GridColor](0.500000,29.000000)(0.500000,32.000000)
\psline[linestyle=dashed,dash=2pt 3pt,dotsep=\GridLineDotSep,linewidth=\GridLineWidth,linecolor=GridColor](0.600000,29.000000)(0.600000,32.000000)
\psline[linestyle=dashed,dash=2pt 3pt,dotsep=\GridLineDotSep,linewidth=\GridLineWidth,linecolor=GridColor](0.700000,29.000000)(0.700000,32.000000)
\psline[linestyle=dashed,dash=2pt 3pt,dotsep=\GridLineDotSep,linewidth=\GridLineWidth,linecolor=GridColor](0.800000,29.000000)(0.800000,32.000000)
\psline[linestyle=dashed,dash=2pt 3pt,dotsep=\GridLineDotSep,linewidth=\GridLineWidth,linecolor=GridColor](0.900000,29.000000)(0.900000,32.000000)
\psline[linestyle=dashed,dash=2pt 3pt,dotsep=\GridLineDotSep,linewidth=\GridLineWidth,linecolor=GridColor](1.000000,29.000000)(1.000000,32.000000)
\psline[linestyle=dashed,dash=2pt 3pt,dotsep=\GridLineDotSep,linewidth=\GridLineWidth,linecolor=GridColor](0.200000,29.000000)(1.000000,29.000000)
\psline[linestyle=dashed,dash=2pt 3pt,dotsep=\GridLineDotSep,linewidth=\GridLineWidth,linecolor=GridColor](0.200000,29.500000)(1.000000,29.500000)
\psline[linestyle=dashed,dash=2pt 3pt,dotsep=\GridLineDotSep,linewidth=\GridLineWidth,linecolor=GridColor](0.200000,30.000000)(1.000000,30.000000)
\psline[linestyle=dashed,dash=2pt 3pt,dotsep=\GridLineDotSep,linewidth=\GridLineWidth,linecolor=GridColor](0.200000,30.500000)(1.000000,30.500000)
\psline[linestyle=dashed,dash=2pt 3pt,dotsep=\GridLineDotSep,linewidth=\GridLineWidth,linecolor=GridColor](0.200000,31.000000)(1.000000,31.000000)
\psline[linestyle=dashed,dash=2pt 3pt,dotsep=\GridLineDotSep,linewidth=\GridLineWidth,linecolor=GridColor](0.200000,31.500000)(1.000000,31.500000)
\psline[linestyle=dashed,dash=2pt 3pt,dotsep=\GridLineDotSep,linewidth=\GridLineWidth,linecolor=GridColor](0.200000,32.000000)(1.000000,32.000000)

\psline[linewidth=\AxesLineWidth,linecolor=GridColor](0.200000,29.000000)(0.200000,29.045000)
\psline[linewidth=\AxesLineWidth,linecolor=GridColor](0.300000,29.000000)(0.300000,29.045000)
\psline[linewidth=\AxesLineWidth,linecolor=GridColor](0.400000,29.000000)(0.400000,29.045000)
\psline[linewidth=\AxesLineWidth,linecolor=GridColor](0.500000,29.000000)(0.500000,29.045000)
\psline[linewidth=\AxesLineWidth,linecolor=GridColor](0.600000,29.000000)(0.600000,29.045000)
\psline[linewidth=\AxesLineWidth,linecolor=GridColor](0.700000,29.000000)(0.700000,29.045000)
\psline[linewidth=\AxesLineWidth,linecolor=GridColor](0.800000,29.000000)(0.800000,29.045000)
\psline[linewidth=\AxesLineWidth,linecolor=GridColor](0.900000,29.000000)(0.900000,29.045000)
\psline[linewidth=\AxesLineWidth,linecolor=GridColor](1.000000,29.000000)(1.000000,29.045000)
\psline[linewidth=\AxesLineWidth,linecolor=GridColor](0.200000,29.000000)(0.209600,29.000000)
\psline[linewidth=\AxesLineWidth,linecolor=GridColor](0.200000,29.500000)(0.209600,29.500000)
\psline[linewidth=\AxesLineWidth,linecolor=GridColor](0.200000,30.000000)(0.209600,30.000000)
\psline[linewidth=\AxesLineWidth,linecolor=GridColor](0.200000,30.500000)(0.209600,30.500000)
\psline[linewidth=\AxesLineWidth,linecolor=GridColor](0.200000,31.000000)(0.209600,31.000000)
\psline[linewidth=\AxesLineWidth,linecolor=GridColor](0.200000,31.500000)(0.209600,31.500000)
\psline[linewidth=\AxesLineWidth,linecolor=GridColor](0.200000,32.000000)(0.209600,32.000000)

{ \footnotesize 
\rput[t](0.300000,28.955000){$0.3$}
\rput[t](0.400000,28.955000){$0.4$}
\rput[t](0.500000,28.955000){$0.5$}
\rput[t](0.600000,28.955000){$0.6$}
\rput[t](0.700000,28.955000){$0.7$}
\rput[t](0.800000,28.955000){$0.8$}
\rput[t](0.900000,28.955000){$0.9$}
\rput[t](1.000000,28.955000){$1$}
\rput[r](0.190400,29.000000){$29$}
\rput[r](0.190400,30.000000){$30$}
\rput[r](0.190400,31.000000){$31$}
\rput[r](0.190400,32.000000){$32$}
} 

\pspolygon[linewidth=\AxesLineWidth](0.200000,29.000000)(1.000000,29.000000)(1.000000,32.000000)(0.200000,32.000000)(0.200000,29.000000)

{ \small 
\rput[b](0.600000,28.375000){
\begin{tabular}{c}
$\gamma$\\
\end{tabular}
}

\rput[t]{90}(0.066667,30.500000){
\begin{tabular}{c}
PSNR [dB]\\
\end{tabular}
}
} 

\newrgbcolor{color538.0049}{0  0  1}
\savedata{\mydata}[{
{0.250000,31.030000},{0.400000,31.500000},{0.500000,31.560000},{0.600000,31.540000},{0.750000,31.440000},
{1.000000,31.160000}
}]
\dataplot[plotstyle=line,linestyle=solid,linewidth=\LineWidth,linecolor=color538.0049]{\mydata}

\end{pspicture}%
	\end{minipage}
	\begin{minipage}{0.32\textwidth}



\providelength{\AxesLineWidth}       \setlength{\AxesLineWidth}{0.5pt}
\providelength{\GridLineWidth}       \setlength{\GridLineWidth}{0.4pt}
\providelength{\GridLineDotSep}      \setlength{\GridLineDotSep}{0.4pt}
\providelength{\MinorGridLineWidth}  \setlength{\MinorGridLineWidth}{0.4pt}
\providelength{\MinorGridLineDotSep} \setlength{\MinorGridLineDotSep}{0.8pt}
\providelength{\plotwidth}           \setlength{\plotwidth}{4.5cm} 
\providelength{\LineWidth}           \setlength{\LineWidth}{0.7pt}
\providelength{\MarkerSize}          \setlength{\MarkerSize}{4pt}
\newrgbcolor{GridColor}{0.8 0.8 0.8}

\psset{xunit=1.428571\plotwidth,yunit=0.266667\plotwidth}
\begin{pspicture}(0.183333,28.375000)(1.000000,32.000000)


\psline[linestyle=dashed,dash=2pt 3pt,dotsep=\GridLineDotSep,linewidth=\GridLineWidth,linecolor=GridColor](0.300000,29.000000)(0.300000,32.000000)
\psline[linestyle=dashed,dash=2pt 3pt,dotsep=\GridLineDotSep,linewidth=\GridLineWidth,linecolor=GridColor](0.400000,29.000000)(0.400000,32.000000)
\psline[linestyle=dashed,dash=2pt 3pt,dotsep=\GridLineDotSep,linewidth=\GridLineWidth,linecolor=GridColor](0.500000,29.000000)(0.500000,32.000000)
\psline[linestyle=dashed,dash=2pt 3pt,dotsep=\GridLineDotSep,linewidth=\GridLineWidth,linecolor=GridColor](0.600000,29.000000)(0.600000,32.000000)
\psline[linestyle=dashed,dash=2pt 3pt,dotsep=\GridLineDotSep,linewidth=\GridLineWidth,linecolor=GridColor](0.700000,29.000000)(0.700000,32.000000)
\psline[linestyle=dashed,dash=2pt 3pt,dotsep=\GridLineDotSep,linewidth=\GridLineWidth,linecolor=GridColor](0.800000,29.000000)(0.800000,32.000000)
\psline[linestyle=dashed,dash=2pt 3pt,dotsep=\GridLineDotSep,linewidth=\GridLineWidth,linecolor=GridColor](0.900000,29.000000)(0.900000,32.000000)
\psline[linestyle=dashed,dash=2pt 3pt,dotsep=\GridLineDotSep,linewidth=\GridLineWidth,linecolor=GridColor](1.000000,29.000000)(1.000000,32.000000)
\psline[linestyle=dashed,dash=2pt 3pt,dotsep=\GridLineDotSep,linewidth=\GridLineWidth,linecolor=GridColor](0.300000,29.000000)(1.000000,29.000000)
\psline[linestyle=dashed,dash=2pt 3pt,dotsep=\GridLineDotSep,linewidth=\GridLineWidth,linecolor=GridColor](0.300000,29.500000)(1.000000,29.500000)
\psline[linestyle=dashed,dash=2pt 3pt,dotsep=\GridLineDotSep,linewidth=\GridLineWidth,linecolor=GridColor](0.300000,30.000000)(1.000000,30.000000)
\psline[linestyle=dashed,dash=2pt 3pt,dotsep=\GridLineDotSep,linewidth=\GridLineWidth,linecolor=GridColor](0.300000,30.500000)(1.000000,30.500000)
\psline[linestyle=dashed,dash=2pt 3pt,dotsep=\GridLineDotSep,linewidth=\GridLineWidth,linecolor=GridColor](0.300000,31.000000)(1.000000,31.000000)
\psline[linestyle=dashed,dash=2pt 3pt,dotsep=\GridLineDotSep,linewidth=\GridLineWidth,linecolor=GridColor](0.300000,31.500000)(1.000000,31.500000)
\psline[linestyle=dashed,dash=2pt 3pt,dotsep=\GridLineDotSep,linewidth=\GridLineWidth,linecolor=GridColor](0.300000,32.000000)(1.000000,32.000000)

\psline[linewidth=\AxesLineWidth,linecolor=GridColor](0.300000,29.000000)(0.300000,29.045000)
\psline[linewidth=\AxesLineWidth,linecolor=GridColor](0.400000,29.000000)(0.400000,29.045000)
\psline[linewidth=\AxesLineWidth,linecolor=GridColor](0.500000,29.000000)(0.500000,29.045000)
\psline[linewidth=\AxesLineWidth,linecolor=GridColor](0.600000,29.000000)(0.600000,29.045000)
\psline[linewidth=\AxesLineWidth,linecolor=GridColor](0.700000,29.000000)(0.700000,29.045000)
\psline[linewidth=\AxesLineWidth,linecolor=GridColor](0.800000,29.000000)(0.800000,29.045000)
\psline[linewidth=\AxesLineWidth,linecolor=GridColor](0.900000,29.000000)(0.900000,29.045000)
\psline[linewidth=\AxesLineWidth,linecolor=GridColor](1.000000,29.000000)(1.000000,29.045000)
\psline[linewidth=\AxesLineWidth,linecolor=GridColor](0.300000,29.000000)(0.308400,29.000000)
\psline[linewidth=\AxesLineWidth,linecolor=GridColor](0.300000,29.500000)(0.308400,29.500000)
\psline[linewidth=\AxesLineWidth,linecolor=GridColor](0.300000,30.000000)(0.308400,30.000000)
\psline[linewidth=\AxesLineWidth,linecolor=GridColor](0.300000,30.500000)(0.308400,30.500000)
\psline[linewidth=\AxesLineWidth,linecolor=GridColor](0.300000,31.000000)(0.308400,31.000000)
\psline[linewidth=\AxesLineWidth,linecolor=GridColor](0.300000,31.500000)(0.308400,31.500000)
\psline[linewidth=\AxesLineWidth,linecolor=GridColor](0.300000,32.000000)(0.308400,32.000000)

{ \footnotesize 
\rput[t](0.400000,28.955000){$0.4$}
\rput[t](0.500000,28.955000){$0.5$}
\rput[t](0.600000,28.955000){$0.6$}
\rput[t](0.700000,28.955000){$0.7$}
\rput[t](0.800000,28.955000){$0.8$}
\rput[t](0.900000,28.955000){$0.9$}
\rput[t](1.000000,28.955000){$1$}
\rput[r](0.291600,29.000000){$29$}
\rput[r](0.291600,30.000000){$30$}
\rput[r](0.291600,31.000000){$31$}
\rput[r](0.291600,32.000000){$32$}
} 

\pspolygon[linewidth=\AxesLineWidth](0.300000,29.000000)(1.000000,29.000000)(1.000000,32.000000)(0.300000,32.000000)(0.300000,29.000000)

{ \small 
\rput[b](0.650000,28.375000){
\begin{tabular}{c}
$\hat{\rho}$\\
\end{tabular}
}

\rput[t]{90}(0.183333,30.500000){
\begin{tabular}{c}
PSNR [dB]\\
\end{tabular}
}
} 

\newrgbcolor{color359.0083}{0  1  0}
\savedata{\mydata}[{
{0.400000,31.370000},{0.500000,31.480000},{0.600000,31.570000},{0.700000,31.560000},{0.800000,31.240000},
{0.900000,30.240000}
}]
\dataplot[plotstyle=line,linestyle=solid,linewidth=\LineWidth,linecolor=color359.0083]{\mydata}

\end{pspicture}%
	\end{minipage}
	\caption{Average reconstruction quality in PSNR for Kodak test data base for varying number of maximum iterations $I$, orthogonality deficiency compensation factors $\gamma$, and weighting function decay factors $\hat{\rho}$.}
	\label{fig:kodak_psnr_over_parameters}
\end{figure*}
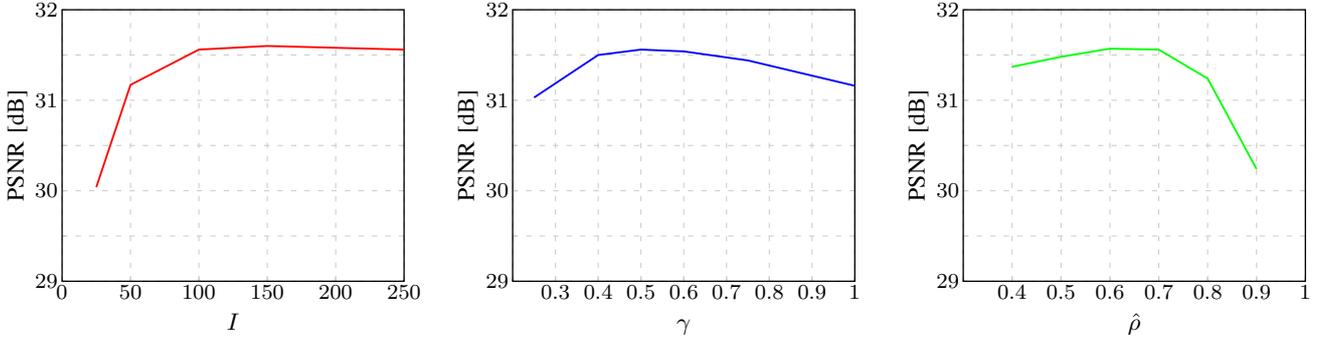

For both three-quarter sampling cases, the same reconstruction algorithms can be used. First, this is again a pixel enlargement (PE) where the output of averaging over the three quadrants just is assigned to all four quadrants. The second method is to apply a modeling by Matching Pursuits (MP)~\cite{Mallat1993} which performs a greedy sparse modeling. The block size and the basis functions for MP are selected as for JSDE. However, unlike JSDE, MP does not include a spatial weighting function, a frequency prior or the orthogonality deficiency compensation. Since MP already is a rather old sparse modeling algorithm, we also include the Generalized Approximate Message Passing (GAMP)~\cite{Rangan2011} as an algorithm for generating a sparse model, given the acquired signal. Finally, the proposed JSDE is applied and evaluated. Since both sensor layouts require a joint deconvolution and extrapolation for reconstructing the image on the high-resolution grid, the novel JSDE can be applied in both scenarios.

The following subsection is devoted to the realization of the JSDE algorithm and especially which parameters to select for the model generation. Afterwards, a comparison of the different sensor concepts and reconstruction algorithms in terms of Peak Signal-to-Noise Ratio (PSNR) and Structural Similarity (SSIM) \cite{Wang2004} follows. This kind of comparison also is used for evaluating the sensitivity of the different sensor layouts. Subsequent to this, a comparison in terms of resolution is provided which shows the superiority of the non-regular sampling concept in contrast to regular sampling. After this, a short subsection follows which discusses the runtime and therewith computational complexity of the algorithms before visual results are provided in the last subsection.

\subsection{Selection of Reconstruction Parameters}

Regarding the JSDE algorithm as it is proposed in Section~\ref{sec:reconstruction}, it can be observed that the algorithm requires several parameters which have to be determined. As mentioned in the preceding section, the model generation of JSDE is related to FSR, even though the latter one only is able to perform an extrapolation and no spatial varying deconvolution. Nevertheless, the parameters given in~\cite{Seiler2015} can be used as a good starting point for determining the parameters for JSDE. As the available image information is uniformly distributed over the whole image area, a data dependent processing order of the blocks as proposed in~\cite{Seiler2015, Seiler2011c} is not required and JSDE can operate in a fixed linescan order of the image blocks.

\begin{table}
	\caption{Model generation parameters for JSDE}
	\begin{center}
		\begin{tabular}{|l|c|}
			\hline Block size $B$ & $4\times 4$ \\ 
			\hline Neighborhood width $W$ & $14$ \\ 
			\hline Iterations $I$ & $100$ \\ 
			\hline Decay factor $\hat{\rho}$ & $0.7$  \\ 
			\hline Orthogonality deficiency compensation $\gamma$ & $0.5$ \\ 
			\hline 
		\end{tabular} 
		\label{tab:reconstruction_parameters}
	\end{center}
\end{table}

As a fitting of the parameters to the underlying data set shall be avoided, the determination of the parameters has to be carried out on a data set which is independent from the later used actual test data set. Hence, we have used the Kodak test data base~\cite{Kodak_Test_Data} for determining the parameters, while for the subsequent evaluation given in the next subsection the TECNICK image data base~\cite{Asuni2011} has been used.

For determining the actual parameter set, a large number of different parameter combinations is evaluated using images from the Kodak test data base. As the proposed sensor design currently only considers the acquisition of the luminance, for the simulations also only the luminance component of the images is considered. Apparently, a full search of the parameter space is not feasible. Hence, we have used the parameters from~\cite{Seiler2015} as starting point and varied them. As metric for the evaluation, the PSNR between the original high-resolution image and the reconstructed image is used.

The simulations on the Kodak image data base reveal that a block size of $B=4$ samples and a border width of $W=14$ samples yields a high reconstruction quality. For the actual model generation, $I=100$ iterations should be carried out and the weighting function should decay with $\hat{\rho}=0.7$. The orthogonality deficiency compensation should be performed with $\gamma=0.5$. In order to provide a compact overview, Table~\ref{tab:reconstruction_parameters} lists all selected parameters.

Fortunately, none of the parameters is very critical and a variation around the selected values is possible without heavily affecting the reconstruction quality. In order to prove this, Figure~\ref{fig:kodak_psnr_over_parameters} shows three plots of the average reconstruction quality where either the number of iterations $I$, the compensation factor $\gamma$, or the decay factor $\hat{\rho}$ are varied while all other parameters are selected according to Table~\ref{tab:reconstruction_parameters}.

\subsection{Evaluation of the Image Quality in Terms of PSNR and SSIM}

Using the above determined parameters, simulations have been carried out on an independent test data set. For this, the TECNICK image data base~\cite{Asuni2011} has been used. The high-resolution images have a size of $1200\times 1200$ pixels. By combining every time $2\times 2$ pixels according to the four considered sensor layouts shown in Figure~\ref{fig:test_sensor_layouts}, all the low-resolution images consist of $600\times 600$ pixels. 

\begin{table}
	\caption{Average image quality in PSNR and SSIM of different sensor layouts and reconstruction strategies for TECNICK image data base~\cite{Asuni2011}.}
	\begin{center}
		\begin{tabular}{|l|l|c|c|}
			\hline 
			Sensor layout	  					& Reconstruction 					& $\PSNR$ 		& SSIM \\
												& algorithm							& in $ \dB$		&\\ \hline 
			Large pixel 						& PE 								& $32.31$   	& $0.9255$ \\ 
												& BIC								& $34.99$	 	& $0.9454$ \\ 
												& SR-Yang~\cite{Yang2010}			& $36.14$		& $0.9540$ \\
												& SR-Kim~\cite{Kim2010a}			& $36.24$		& $0.9550$ \\
												& SR-Zeyde~\cite{Zeyde2010}			& $36.44$		& $0.9568$ \\  \hline
			Non-regular quarter 				& FSR~\cite{Seiler2015} 			& $34.79$	 	& $0.9408$ \\ 
				sampling~\cite{Schoeberl2011}	& KR~\cite{Takeda2007}				& $33.36$	 	& $0.9204$ \\ 
												& CLS~\cite{Afonso2011}				& $32.10$	 	& $0.9116$ \\ \hline 
			Regular three-quarter 				& PE					 			& $31.85$	 	& $0.9199$ \\ 
				sampling						& MP~\cite{Mallat1993}				& $\hphantom{0}7.35$ 		& $0.0391$\\ 
												& GAMP~\cite{Rangan2011}			& $\hphantom{0}5.39$ 		& $0.0038$ \\ 
												& JSDE								& $35.16$	 	& $0.9438$ \\ \hline 
			Non-regular three-					& PE				 				& $31.86$	 	& $0.9194$ \\ 
				quarter sampling				& MP~\cite{Mallat1993}				& $26.80$	 	& $0.7111$ \\ 
												& GAMP~\cite{Rangan2011}			& $27.16$ 		& $0.7434$ \\ 
												& JSDE								& $35.57$	 	& $0.9474$ \\ \hline 
		\end{tabular} 
		\label{tab:reconstruction_evaluation}
	\end{center}
\end{table}

In Table~\ref{tab:reconstruction_evaluation}, the average image quality in terms of PSNR and SSIM is listed for the TECNICK image data base for the above mentioned sensor layouts and the considered reconstruction strategies. It can be seen that the proposed non-regular three-quarter sampling in combination with the novel JSDE reconstruction yields a very high objective image quality, outperforming all other combinations except for the SR algorithms. 

Especially compared to the case that a large pixel is considered in combination with bicubic interpolation, a gain of more than $0.5 \dB$ can be achieved. Comparing the proposed non-regular three-quarter sampling to the non-regular quarter sampling~\cite{Schoeberl2011}, it can be observed that by applying JSDE for the reconstruction, a gain of more than $0.7 \dB$ is possible, and at the same time three times the light is collected, compared to the quarter sampling case.

Regarding the reconstruction output of JSDE for the case that the three-quarter sampling is carried out in a regular or in a non-regular fashion, it can be observed that the non-regular rotation of the pixel cell yields a gain of more than $0.4 \dB$. This is due to the fact that the non-regular placement of the light-sensitive regions does not produce the typical aliasing components and allows for the reconstruction of even very fine details. Looking at the results for MP and GAMP, it can be seen that the latter two only produce a significantly lower quality than JSDE and especially for the regular case completely fail. This can be explained by the fact that MP and GAMP have no frequency prior included. Thus, for the regular sampling case these algorithms are not able to distinguish between the selection of low-frequency and high-frequency basis functions and the actual selection process is determined by numerical inaccuracies. Hence, MP and GAMP produce a lot of artifacts in this special case, resulting in a very low quality. In this context, it always has to be kept in mind, that a better approximation of the available signal does not necessarily have to come along with a higher reconstruction quality. Accordingly, as MP and GAMP do not employ the spatial weighting function, the frequency prior, and the orthogonality deficiency compensation, they approximate the available signal better than JSDE. However, they are not able to achieve a higher reconstruction quality. Another interesting aspect which can also be discovered is that PE already allows for a decent reconstruction quality for the three-quarter sampling cases. Thus, it would also be possible to apply an elaborate reconstruction as by JSDE offline, while for a preview just PE is used. 

Even though the non-regular three-quarter sampling achieves a high reconstruction quality, it is still outperformed in PSNR and SSIM if the SR algorithms are applied on the large pixel case. As outlined above, single-image SR algorithms are able to achieve a high quality by exploiting the self-similarity in images or similarities to trained data sets. Since the training has been carried out on images with a similar characteristic as the test data, the output of the considered algorithms is rather high. However, in all the cases when content is considered that does not fulfill these properties, they fail. And more important, even though the SR algorithms can improve the PSNR, they are not able to increase the actual resolution of an imaging system. The spatial resolution always is limited by the sampling in the image sensor and as shown in the subsection after the next, cannot be improved by single-image SR algorithms.

\begin{table}
	\caption{Average image quality in PSNR and SSIM of different sensor layouts and reconstruction strategies for TECNICK image data base~\cite{Asuni2011} in case of superimposed noise. The loss compared to noiseless is given in brackets.}
	\begin{center}
		\begin{tabular}{|l|l|c|c|}
			\hline 
			Sensor layout	  				& Reconstruction 					& $\PSNR$  					& SSIM  \\
											& algorithm							& in $ \dB$					&  \\ \hline 
			Large pixel 					& PE 								& $32.17\ (0.15)$	 		& $0.9177\ (0.0088)$ \\ 
											& BIC								& $34.79\ (0.19)$ 			& $0.9406\ (0.0048)$ \\ 
											& SR-Yang~\cite{Yang2010}			& $35.49\ (0.64)$			& $0.9397\ (0.0142)$\\
											& SR-Kim~\cite{Kim2010a}			& $35.71\ (0.53)$			& $0.9459\ (0.0107)$\\
											& SR-Zeyde~\cite{Zeyde2010}			& $35.89\ (0.55)$			& $0.9459\ (0.0109)$\\  \hline
			Non-regular			 			& FSR~\cite{Seiler2015}				& $34.16\ (0.63)$  			& $0.9231\ (0.0176)$ \\ 
			quarter 						& KR~\cite{Takeda2007}				& $33.05\ (0.31)$  			& $0.9105\ (0.0099)$ \\ 
			sampling~\cite{Schoeberl2011}	& CLS~\cite{Afonso2011}				& $31.83\ (0.27)$  			& $0.8994\ (0.0122)$\\ \hline 
			Regular 					 	& PE					 			& $31.68\ (0.17)$  			& $0.9097\ (0.0102)$ \\ 
			three-quarter					& MP~\cite{Mallat1993}				& $\hphantom{0}7.22\ (0.14)$& $0.0284\ (0.0107)$ \\ 
			sampling						& GAMP~\cite{Rangan2011}			& $\hphantom{0}5.39\ (0.00)$			& $0.0038\ (0.0000)$	\\ 
											& JSDE								& $34.84\ (0.31)$  			& $0.9351\ (0.0087)$ \\ \hline 
			Non-regular  					& PE				 				& $31.68\ (0.17)$  			& $0.9092\ (0.0102)$\\ 
			three-quarter					& MP~\cite{Mallat1993}				& $26.59\ (0.21)$  			& $0.6831\ (0.0280)$\\ 
			sampling						& GAMP~\cite{Rangan2011}			& $27.16\ (0.00)$ 			& $0.7433\ (0.0001)$\\ 
											& JSDE								& $35.23\ (0.34)$  			& $0.9389\ (0.0084)$ \\ \hline 
		\end{tabular} 
		\label{tab:reconstruction_evaluation_noise}
	\end{center}
\end{table}

\begin{figure*}
	\centering
	\psfrag{Original}[l][l][0.75]{Original}
	\psfrag{LP+BiCubic}[l][l][0.75]{LP + BIC}
	\psfrag{LP+SR-Zeyde}[l][l][0.75]{LP + SR-Zeyde}
	\psfrag{1/4-NonReg+FSR}[l][l][0.75]{$\frac{1}{4}$-NonReg + FSR}
	\psfrag{1/4-NonReg+KR}[l][l][0.75]{$\frac{1}{4}$-NonReg + KR}
	\psfrag{3/4-Reg+Enlargement}[l][l][0.75]{$\frac{3}{4}$-Reg + PE}
	\psfrag{3/4-Reg+JSDE}[l][l][0.75]{$\frac{3}{4}$-Reg + JSDE}
	\psfrag{3/4-NonReg+Enlargement}[l][l][0.75]{$\frac{3}{4}$-NonReg + PE}
	\psfrag{3/4-NonReg+JSDE}[l][l][0.75]{$\frac{3}{4}$-NonReg + JSDE}
	\includegraphics[width=0.95\textwidth]{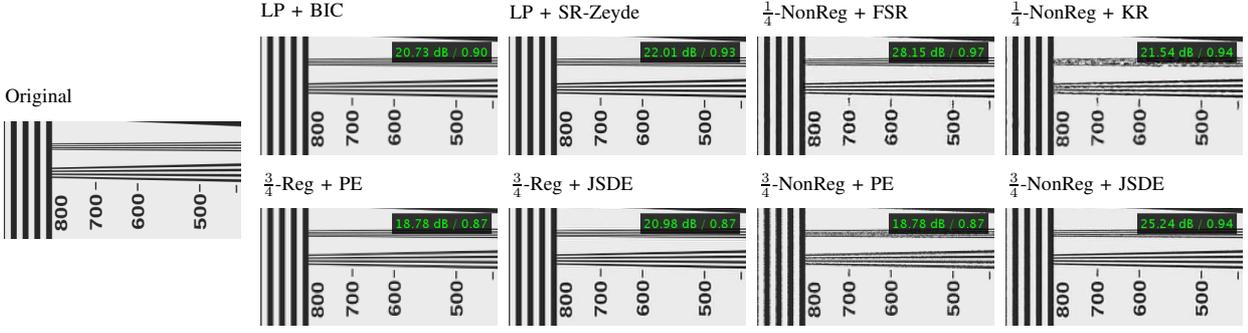}
	\caption{Reconstruction results for small area of the EIA-1956 Resolution Test Chart. Considered sensor layouts are large pixels (LP), non-regular quarter sampling ($\frac{1}{4}$-NonReg), regular three-quarter sampling ($\frac{3}{4}$-Reg), and non-regular three-quarter sampling ($\frac{3}{4}$-NonReg). In the top right border, PSNR and SSIM results are given. \emph{(Please pay attention, additional aliasing may be caused by printing or scaling. Best to be viewed enlarged on a monitor.)} }
	\label{fig:results_EIA_testchart}
\end{figure*}

\begin{figure} 
	\centering
	\input{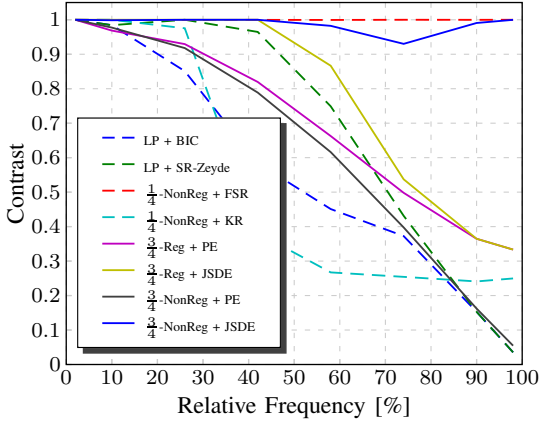}
	\caption{Modulation transfer function showing the contrast with respect to the spatial frequency for different sensor layouts and reconstruction algorithms. The frequency of the line pattern is defined relative to the sampling frequency.}
	\label{fig:mtf_evaulation}
\end{figure}

\subsection{Evaluation of the Sensitivity}
\label{ssec:sensitivity}

Independent of the three-quarter sampling being achieved by placing the insensitive parts of a pixel cell in one quadrant or by using a mask in front of the pixels, it has to be considered that a fill factor of only $75\%$ can be used. In contrast to this, common image sensors with backside illumination and / or microlenses can achieve fill factors of up to roughly $100\%$. Accordingly, the proposed three-quarter sampling has a lower sensitivity and hence is more prone to shot-noise. In order to assess the influence of the reduced fill factor of the proposed concept on the actual imaging quality, simulations with superimposed noise have been carried out. For this, shot noise following a Poisson distribution has been added to the pixels in order to simulate the variation of the incident light. In order to simulate the noise, a full well capacity of $10.000 \mathrm{e}^-$ has been assumed. Compared to state-of-the-art imaging sensors, this is a quite low full well capacity, however, in order to actually assess the effects, this more challenging scenario has been considered. Additionally, in order to simulate thermal and readout noise, Gaussian distributed noise with a standard deviation of $25 \mathrm{e}^-$ has been added. Table \ref{tab:reconstruction_evaluation_noise} lists the average image quality in PSNR and SSIM for the TECNICK image data base, again. For this, the above mentioned sensor layouts and the considered reconstruction strategies are employed. Apparently, the image quality is lower compared to the noiseless case. Furthermore, it can be observed that in the case where the large pixels are used, the loss is very small, whereas for non-regular quarter sampling \cite{Schoeberl2011}, the highest loss arises. The large loss of the non-regular quarter sampling is caused by the small area sensitive to light. Since the proposed non-regular three-quarter sampling integrates the light over a larger area, it is less sensitive to noise and degradation caused by the noise is moderate.

\subsection{Evaluation of the Image Resolution}
\label{ssec:image_resolution}

\begin{figure*}
	\centering
	\psfrag{Original}[l][l][0.75]{Original}
	\psfrag{LP+BiCubic}[l][l][0.75]{LP + BIC}
	\psfrag{LP+SR-Zeyde}[l][l][0.75]{LP + SR-Zeyde}
	\psfrag{1/4-NonReg+FSR}[l][l][0.75]{$\frac{1}{4}$-NonReg + FSR}
	\psfrag{1/4-NonReg+KR}[l][l][0.75]{$\frac{1}{4}$-NonReg + KR}
	\psfrag{3/4-Reg+Enlargement}[l][l][0.75]{$\frac{3}{4}$-Reg + PE}
	\psfrag{3/4-Reg+JSDE}[l][l][0.75]{$\frac{3}{4}$-Reg + JSDE}
	\psfrag{3/4-NonReg+Enlargement}[l][l][0.75]{$\frac{3}{4}$-NonReg + PE}
	\psfrag{3/4-NonReg+JSDE}[l][l][0.75]{$\frac{3}{4}$-NonReg + JSDE}
	\includegraphics[width=0.8\textwidth]{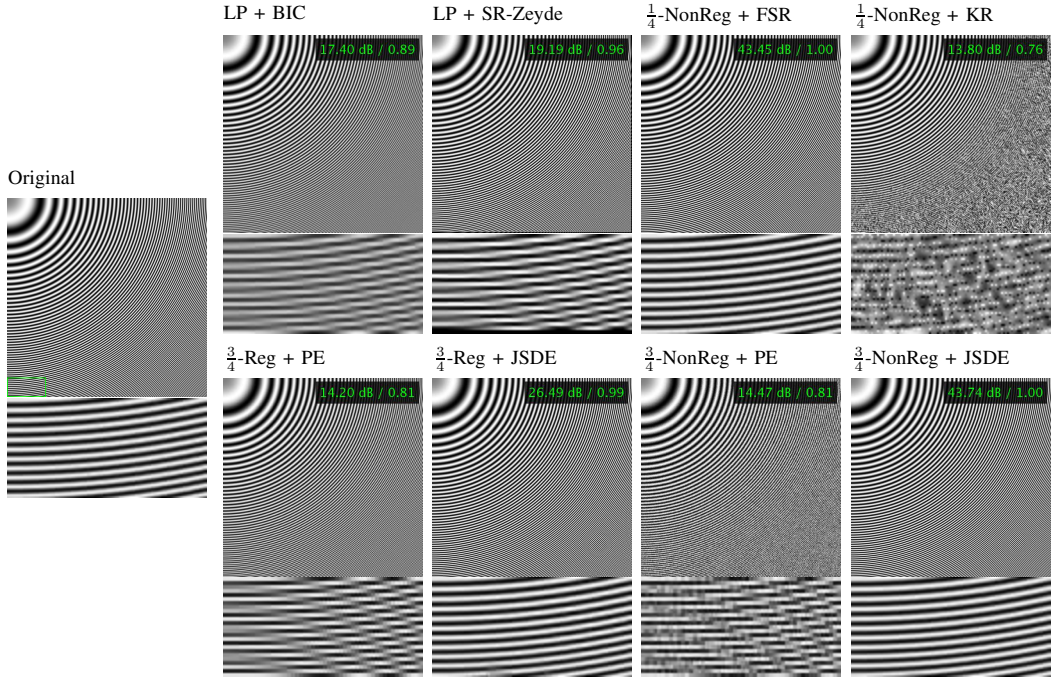}
	\caption{Visual results for one quarter of test image Zoneplate. Detail areas from the location marked in green in the left most image are provided below each image. Results are provided for sensor layouts large pixels (LP), non-regular quarter sampling ($\frac{1}{4}$-NonReg), regular three-quarter sampling ($\frac{3}{4}$-Reg), and non-regular three-quarter sampling ($\frac{3}{4}$-NonReg) and considered reconstruction algorithms. In the top right border, PSNR and SSIM results are given. \emph{(Please pay attention, additional aliasing may be caused by printing or scaling. Best to be viewed enlarged on a monitor.)} }
	\label{fig:visual_results_zoneplate}
\end{figure*}

The evaluation of the quality in terms of PSNR and SSIM is only one possibility to measure the image quality. Aside from this, the achievable image resolution has to be evaluated. PSNR and SSIM only provide an overall quality score but do not express the abilities of a system to resolve even very fine details. In order to assess this property, different evaluations have been carried out and are discussed in the following. In order to allow for a compact presentation, only a subset of the aforementioned combinations of sensor layouts and good performing reconstruction algorithms is considered in this subsection. For obtaining a representative set of combinations, the large pixel case is considered together with BIC and SR-Zeyde~\cite{Zeyde2010}, the non-regular quarter sampling together with FSR~\cite{Seiler2015} and KR~\cite{Takeda2007}, and for the regular and non-regular three-quarter sampling, PE and the proposed JSDE is considered. 

The first option to evaluate the image resolution is to apply the different combinations on a resolution test chart and determine how well different details can be recognized. For this, the EIA-1956 Resolution Test Chart is considered. In Figure~\ref{fig:results_EIA_testchart}, details of the reconstructed images are shown. It can be observed that all regular sensor layouts lead to severe aliasing and that small details cannot be separated well. Only the non-regular quarter and three-quarter sampling in combination with FSR and the proposed JSDE are able to acquire even fine details and separate the converging lines up to the smallest distance. In direct comparison between the quarter sampling and the three-quarter sampling case, it can be discovered that FSR achieves a higher resolution and that the converging lines can be separated more precisely. In doing so, also a higher PSNR can be achieved. However, the non-regular quarter sampling in combination with FSR also introduces some ringing and some ragged edges at the letters. Nevertheless, the three-quarter sampling in combination with JSDE also is able to resolve even fine details and does not produce these artifacts.

In order to quantify the abilities of the different sampling concepts and reconstruction algorithms, further tests have been carried out with line patterns of different spatial frequency. For measuring the resolution of the different combinations, the modulation transfer function is considered that measures the contrast of the reconstructed signal for different spatial frequencies. The contrast $C$ is defined as
\begin{equation}
C= \frac{I_\mathrm{max} - I_\mathrm{min}}{I_\mathrm{max}+I_\mathrm{min}}
\end{equation}
with $I_\mathrm{max}$ being the maximum amplitude in the reconstructed image and $I_\mathrm{min}$ being the minimum amplitude. In Figure~\ref{fig:mtf_evaulation}, the contrast is plotted with respect to the frequency of the line pattern, defined relative to the spatial sampling frequency with respect to the low-resolution grid. It can be seen that both non-regular sampling layouts are able to achieve a high contrast up to the sampling frequency of the low-resolution grid if combined with FSR, or respectively, JSDE. The non-regular quarter sampling achieves an almost perfect reconstruction in this case. Nevertheless, the non-regular three-quarter sampling also is able to yield a very high contrast and to resolve even very fine structures. Of course, due to aliasing the regular sampling with large pixels can only cover frequencies up to half the sampling frequency and the contrast drops significantly if the Nyquist-frequency is exceeded. This also cannot be resolved by single image SR algorithms and it can be observed that in this case the resolution is limited by the underlying large pixels. For high spatial frequencies, the SR algorithm can only achieve a resolution similar to the BIC case.

In order to show that the ability to resolve fine details is not direction-dependent, Figure~\ref{fig:visual_results_zoneplate} shows details of the test image Zoneplate which is a rotation-symmetric chirp and is shown at the left side. This test image covers the whole frequency range in all directions and is well suited for measuring to what extent high-frequency content can be recovered. Comparing the images, it can be discovered that only the non-regular quarter sampling in combination with FSR and the proposed non-regular three-quarter sampling in combination with JSDE are able to cover the whole frequency range and acquire even very fine details. All other algorithms produce strong aliasing artifacts for high frequencies, also leading to a significant loss in PSNR and SSIM. Regarding the case of regular three-quarter sampling in combination with JSDE, it can be seen that aliasing artifacts occur there, as well. However, they are not an output of the reconstruction process, but rather result from the regular sampling pattern.

\subsection{Evaluation of the Computational Complexity}

\begin{table}
	\caption{Average processing time for reconstructing one image of size $1024\times 1024$ pixel.}
	\begin{center}
		\begin{tabular}{|l|l|r|}
			\hline 
			Sensor layout	  					& Reconstruction 					& \multicolumn{1}{l|}{Processing time}  \\
												& algorithm							& \multicolumn{1}{l|}{in sec}				 \\ \hline 
			Large pixel 						& PE 								& $0.0067$ 		 \\ 
												& BIC								& $0.016$ 		 \\ 
												& SR-Yang~\cite{Yang2010}			& $734.99$				 \\
												& SR-Kim~\cite{Kim2010a}			& $557.98$				 \\
												& SR-Zeyde~\cite{Zeyde2010}			& $20.89$					 \\  \hline
			Non-regular quarter 				& FSR~\cite{Seiler2015}				& $208.78$  		 \\ 
			sampling~\cite{Schoeberl2011}		& KR~\cite{Takeda2007}				& $137.16$  		 \\ 
												& CLS~\cite{Afonso2011}				& $52.84$  		\\ \hline 
			Regular three-quarter			 	& PE					 			& $0.014$  		 \\ 
			sampling							& MP~\cite{Mallat1993}				& $19794.67$   \\ 
												& GAMP~\cite{Rangan2011}			& $13142.46$ 					 	\\ 
												& JSDE								& $20524.30$  		\\ \hline 
			Non-regular three- 					& PE				 				& $0.44$  		\\ 
			quarter sampling					& MP~\cite{Mallat1993}				& $20464.58$  		\\ 
												& GAMP~\cite{Rangan2011}			& $8092.05$ 					 \\ 
												& JSDE								& $20523.00$  		 \\ \hline 
		\end{tabular} 
		\label{tab:runtime_evaluation}
	\end{center}
\end{table}

For evaluating the computational complexity of the different reconstruction algorithms, runtime tests have been carried out with MATLAB R2016b on an Intel Xeon E5-1620 v2 equipped with 32 GB RAM. As some of the reconstruction algorithms make use of MEX-Files, the comparison is not completely fair, nevertheless it provides a good impression of the overall complexity. In order to get reliable results, $10$ runs have been carried out and always the same image of size $1024\times 1024$ pixels has been reconstructed.

\begin{figure*}
	\centering
	\psfrag{Original}[l][l][1]{Original}
	\psfrag{LP+BiCubic}[l][l][1]{LP + BIC}
	\psfrag{LP+SR-Zeyde}[l][l][1]{LP + SR-Zeyde}
	\psfrag{1/4-NonReg+FSR}[l][l][1]{$\frac{1}{4}$-NonReg + FSR}
	\psfrag{1/4-NonReg+KR}[l][l][1]{$\frac{1}{4}$-NonReg + KR}
	\psfrag{3/4-Reg+Enlargement}[l][l][1]{$\frac{3}{4}$-Reg + PE}
	\psfrag{3/4-Reg+JSDE}[l][l][1]{$\frac{3}{4}$-Reg + JSDE}
	\psfrag{3/4-NonReg+Enlargement}[l][l][1]{$\frac{3}{4}$-NonReg + PE}
	\psfrag{3/4-NonReg+JSDE}[l][l][1]{$\frac{3}{4}$-NonReg + JSDE}
	\includegraphics[width=0.99\textwidth]{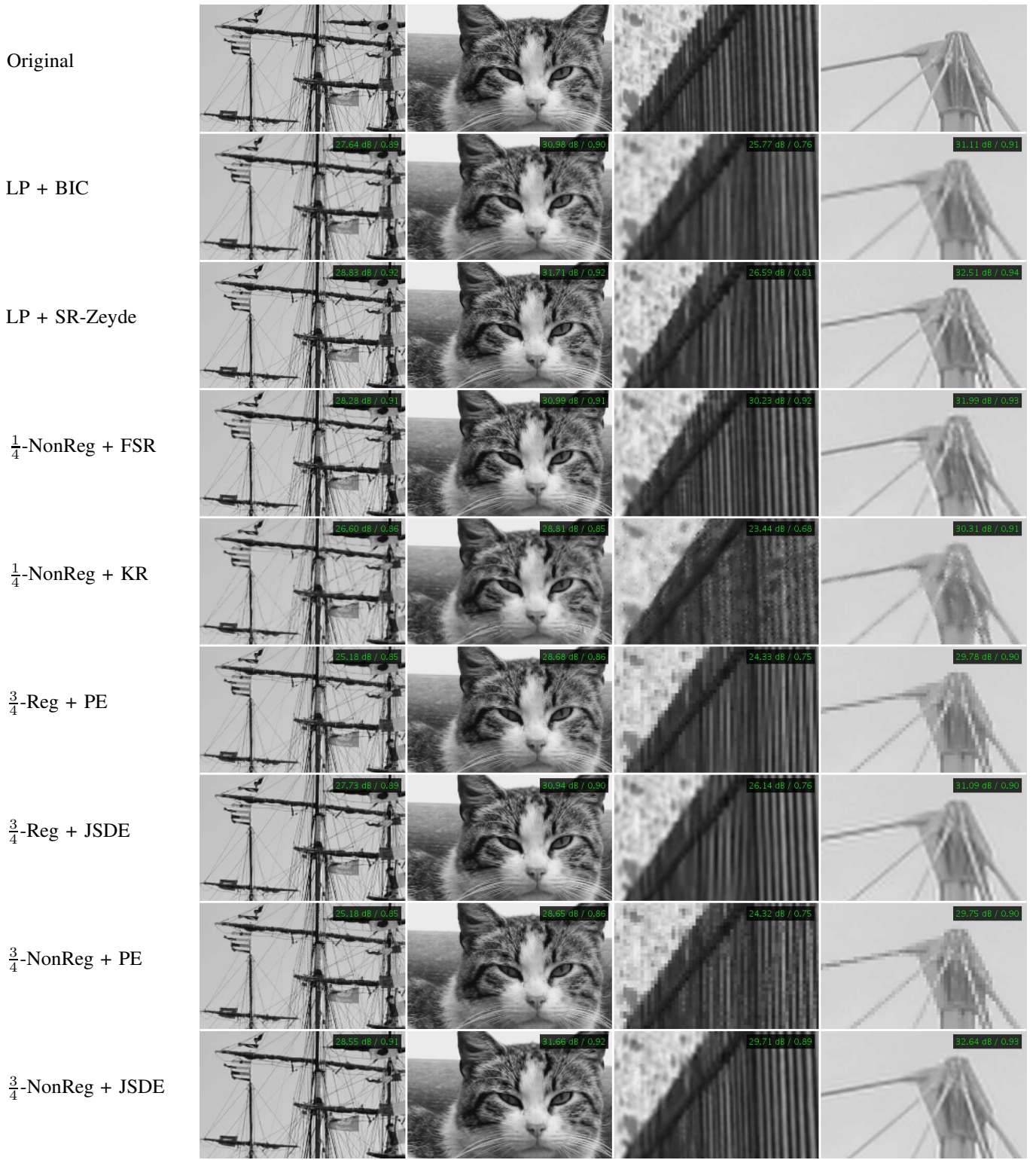}
	\caption{Visual results for details of $400\times 250$ pixels (left two columns) and $120\times 75$ pixels (right two columns) of different test images from the TECNICK image data base~\cite{Asuni2011}. The sensor layouts large pixels (LP), non-regular quarter sampling ($\frac{1}{4}$-NonReg), regular three-quarter sampling ($\frac{3}{4}$-Reg), and non-regular three-quarter sampling ($\frac{3}{4}$-NonReg) are considered in combination with different reconstruction algorithms. \emph{(Please pay attention, additional aliasing may be caused by printing or scaling. Best to be viewed enlarged on a monitor.)
		} }
		\label{fig:visual_results}
	\end{figure*}

In Table~\ref{tab:runtime_evaluation}, the average runtime in seconds is provided for the different combinations of sensor layouts and reconstruction algorithms. It can be observed, that the runtime spreads over several magnitudes from simple reconstruction algorithms like PE and BIC to the proposed JSDE. Currently, JSDE is the most complex algorithm even though its complexity is of the same order as the other sparse modeling algorithms MP and GAMP. Even though the runtime of JSDE looks poor at first glance, the algorithm has potential for high speed-ups. For this, the relationship of JSDE to FSR could be exploited and a frequency domain implementation of JSDE similar to FSR \cite{Seiler2015} is envisaged for the future. Using this strategy, as shown in \cite{Genser2017}, for FSR even a real-time operation of FSR is possible.

\subsection{Evaluation of the Visual Quality}

Aside from the objective evaluation in terms of PSNR/SSIM and resolution provided above, of course, the visual quality of the reconstructed images is of high importance. For demonstrating the visual quality of the different concepts, Figure~\ref{fig:visual_results} shows the output for different combinations of sensor layouts and reconstruction algorithms. In order to allow for a compact presentation, only the same subset of combinations between sensor layouts and reconstruction algorithms as used in Subsection \ref{ssec:image_resolution} is considered, again. The image details have been selected in order to represent very different content and it can be seen that the considered algorithms behave differently with respect to the content to be reconstructed. Furthermore, for each image, the PSNR and the SSIM is provided. However, in many cases the measured PSNR does neither fit well the actual resolution of the image, nor the visual impression.

Comparing the different images, it can be observed that except for the non-regular sampling and reconstructions by FSR and JSDE, all sensor layout and reconstruction algorithm pairs suffer either from aliasing or from introduced artifacts. Looking at the output from the large pixel and the SR algorithm, it can be discovered that this combination yields a high quality in many cases. However, whenever it comes to very fine details, this combination suffers from aliasing, leading to a reduced visual quality. If the non-regular quarter sampling case with FSR reconstruction and the proposed non-regular three-quarter sampling in combination with JSDE are examined in detail, it can be observed that FSR introduces some very small ringing artifacts which are not visible for the JSDE case. Hence, the proposed concept is more sensitive to light and also yields a higher visual image quality.


\section{Conclusion and Outlook} \label{sec:conclusion} 
In this paper, we have proposed a novel strategy for modifying image sensors in order to increase the imaging resolution. By applying a non-regular sampling, the limitations of image sensors caused by aliasing can be avoided. The proposed sensor layout is still based on state-of-the-art technologies and only very small changes of the design tool chain and workflow are required. This is achieved by defining a prototype pixel cell which is non-regularly rotated. Using this, the light falling on the sensor pixels is integrated over non-regularly placed regions, therewith achieving a non-regular sampling. In order to achieve a higher resolution with the proposed sensor, a reconstruction to a high-resolution grid is required. For this, we have proposed the Joint Sparse Deconvolution and Extrapolation (JSDE) as a sparsity-based algorithm which allows for solving the underdetermined problem that comes along with the reconstruction. By performing a spatially varying deconvolution in combination with an extrapolation, JSDE is able to achieve a very high image quality. Using this, non-regular sampling in combination with JSDE is able to outperform classical sensor layouts and reconstruction algorithms in terms of resolution and achieves a high visual quality. 

Future research aims at combining the proposed non-regular sensor layout with super-resolution techniques. As the non-regular sampling is directly able to acquire more high-frequency information than regular image sensors, this might also be beneficial for reconstruction algorithms which are more data-driven than the generic JSDE. Furthermore, the extension of the non-regular sampling concept to color imaging will be investigated. For this, on the one hand a direct combination of the proposed sensor concept with a Bayer pattern can be considered. On the other hand, an extension of the non-regular sampling concept to the placement of the color filters is foreseen. Finally, the manufacturing of a non-regular sampling sensor is envisioned. With such a sensor, actual measurements would be possible, proving the potential of the proposed concept in real-world applications.


\appendix
In the following, the steps that are required for determining the index $u^{\left(\nu\right)}$ of the basis function to add in (\ref{eq:bf_selection}) are outlined in detailed. For this (\ref{eq:bf_selection_origin}) is reformulated as
\[
u^{\left(\nu\right)} = \argmin_k  \bigg(q_k\left(\ve{r}^{\left(\nu-1\right)} - p_k^{\left(\nu\right)} \ve{D}\ve{A}\ve{\varphi}_k\right)^\herm \ve{W} \hspace{1cm}
\]
\begin{equation}
\label{eq:app1} \hspace{4cm} \left(\ve{r}^{\left(\nu-1\right)} - p_k^{\left(\nu\right)} \ve{D}\ve{A}\ve{\varphi}_k\right)\bigg).
\end{equation}
By distributing the product, one obtains
\[
u^{\left(\nu\right)} = \argmin_k \bigg(q_k\left(\ve{r}^{\left(\nu-1\right)} - p_k^{\left(\nu\right)} \ve{D}\ve{A}\ve{\varphi}_k\right)^\herm \ve{W} \ve{r}^{\left(\nu-1\right)} - \hspace{0.5cm}
\]
\begin{equation}
\label{eq:app2} \hspace{1.4cm}  q_k p_k^{\left(\nu\right)} \left(\ve{r}^{\left(\nu-1\right)} - p_k^{\left(\nu\right)} \ve{D}\ve{A}\ve{\varphi}_k\right)^\herm \ve{W}\ve{D}\ve{A}\ve{\varphi}_k \bigg).
\end{equation}
Looking at the second product within the brackets, one can observe that
\[
\left(\ve{r}^{\left(\nu-1\right)} - p_k^{\left(\nu\right)} \ve{D}\ve{A}\ve{\varphi}_k\right)^\herm \ve{W}\ve{D}\ve{A}\ve{\varphi}_k = \hspace{3cm}
\]
\begin{eqnarray*}
	&=& \hspace{-2mm}	\left(\ve{r}^{\left(\nu-1\right)} \hspace{-1mm}- \frac{\left(\ve{D}\ve{A}\ve{\varphi}_k\right)^\herm\ve{W}\ve{r}^{\left(\nu-1\right)}}{\left(\ve{D}\ve{A}\ve{\varphi}_k\right)^\herm\ve{W}\ve{D}\ve{A}\ve{\varphi}_k} \ve{D}\ve{A}\ve{\varphi}_k\right)^\herm \ve{W}\ve{D}\ve{A}\ve{\varphi}_k \\
	&=& \ve{r}^{\left(\nu-1\right)^\herm} \ve{W}\ve{D}\ve{A}\ve{\varphi}_k - \left(\left(\ve{D}\ve{A}\ve{\varphi}_k\right)^\herm\ve{W}\ve{r}^{\left(\nu-1\right)}\right)^\ast \\
	&=& 0, \hspace{10mm} \mbox{since $\ve{W}$ is diagonal and real}.
\end{eqnarray*}
Using this, (\ref{eq:app2}) can be simplified to
\begin{equation}
\label{eq:app3}
u^{\left(\nu\right)} = \argmin_k \left(q_k\left(\ve{r}^{\left(\nu-1\right)} - p_k^{\left(\nu\right)} \ve{D}\ve{A}\ve{\varphi}_k\right)^\herm \ve{W} \ve{r}^{\left(\nu-1\right)}\right).
\end{equation}
By removing the terms which are independent of $k$ and changing the minimization over a negative term to a maximization of the corresponding term with positive sign, one obtains
\begin{equation}
\label{eq:app4}
u^{\left(\nu\right)} = \argmax_k \left(q_k p_k^{\left(\nu\right)^\ast} \left( \ve{D}\ve{A}\ve{\varphi}_k\right)^\herm \ve{W} \ve{r}^{\left(\nu-1\right)}\right).
\end{equation}
Using that 
\begin{equation}
\left( \ve{D}\ve{A}\ve{\varphi}_k\right)^\herm \ve{W} \ve{r}^{\left(\nu-1\right)} =p_k^{\left(\nu\right)}\left( \ve{D}\ve{A}\ve{\varphi}_k\right)^\herm \ve{W}\ve{D}\ve{A}\ve{\varphi}_k 
\end{equation} holds, (\ref{eq:app1}) can be simplified to 
\begin{eqnarray}
\nonumber u^{\left(\nu\right)}  &=& \argmax_k \left( q_k\left| p_k^{\left(\nu\right)}\right|^2\left( \ve{D}\ve{A}\ve{\varphi}_k\right)^\herm \ve{W}\ve{D}\ve{A}\ve{\varphi}_k \right) \\
&=& \argmax_k \left(q_k \frac{\left|\left(\ve{D}\ve{A}\ve{\varphi}_k\right)^\herm\ve{W}\ve{r}^{\left(\nu-1\right)}\right|^2}{\left(\ve{D}\ve{A}\ve{\varphi}_k\right)^\herm\ve{W}\ve{D}\ve{A}\ve{\varphi}_k}\right).
\end{eqnarray}



\end{document}